\documentclass[twocolumn,pra,showpacs,amsmath,amssymb]{revtex4}
\usepackage[dvipdfmx]{graphicx}
\usepackage{amssymb,amsthm,amsfonts,amstext}
\newcommand{\B}{\mathfrak{B}}

\newcommand{\Hi}{\mathcal{H}}

\newcommand{\Tr}{\mathrm{Tr}}
\newcommand{\argmax}{\mathop{\rm argmax}\limits}
\newcommand{\argmin}{\mathop{\rm argmin}\limits}

\newcommand{\ket}[1]{| #1 \rangle}
\newcommand{\bra}[1]{\langle #1 |}
\newcommand{\braket}[2]{\langle #1 \vert #2 \rangle}
\newcommand{\be}{\begin{equation}}
\newcommand{\ee}{\end{equation}}
\newtheorem{Theorem}{Theorem}

\newtheorem{Lemma}{Lemma}
\newtheorem{Corollary}{Corollary}

\usepackage{amsmath}	% required for `\align' (yatex added)
\begin{document}

\title{Asymptotic local hypothesis testing between a pure bipartite state and the
completely mixed state}
\author{Masaki Owari $^{1,2}$ 
        Masahito Hayashi,$^{3,4,5}$}
\address{$^1${\it NTT Communication Science Laboratories, NTT Corporation
3-1,Morinosato Wakamiya Atsugi-Shi, Kanagawa, 243-0198, Japan}\\
$^2${\it Institut f\"ur Theoretische Physik,
        Universit\"at Ulm, Albert-Einstein-Allee 11, D - 89069 Ulm, Germany}\\
$^3${\it  Graduate School of Mathematics,  
Nagoya University, Furocho, Chikusa-ku, Nagoya 464-8602, Japan}\\
$^4${\it  Division of Mathematics, Graduate School of Information Sciences 
Tohoku University, 6-3-09 Aramaki-Aza-Aoba, Aoba-ku, Sendai 980-8579, Japan}\\
$^5${\it  Centre for Quantum Technologies
National University of Singapore, 3 Science Drive 2, 117542, Singapore}
}

\begin{abstract}
In this paper, we treat an asymptotic hypothesis testing (or state 
 discrimination with asymmetric treatment of errors) between an
 arbitrary fixed bipartite pure state $\ket{\Psi}$ and the completely
 mixed state by one-way LOCC, two-way LOCC, and separable POVMs.
As a result, we derive single-letterized formulas for the Stein's lemma type of
optimal error exponents under one-way LOCC, two-way LOCC and separable POVMs, 
the Chernoff bounds  under
one-way LOCC POVMs and separable POVMs, and the Hoeffding bounds under
one-way LOCC POVMs in the whole region of a parameter and  under separable
POVMs on a restricted region of a parameter. We also numerically calculate 
the Chernoff and the Hoeffding bounds under a class of three-step LOCC protocols  in low-dimensional
      systems and show that these bounds not only
      outperform the bounds for one-way LOCC POVMs but also almost approximates the
      bounds for separable POVMs in the parameter region where
      analytical bounds for separable POVMs are derived. 
\end{abstract}

\maketitle

\section{Introduction}
\subsection{Local hypothesis testing problems}
Local discrimination, which is the problem of discriminating an unknown quantum state
from known candidates by means of local operation and classical
communication (LOCC), has been intensively studied since almost the 
advent of the field of quantum information \cite{PW91,MP95,BDFMRSSW99,
WSHV00,GV01,VSPM01,GKRSS01,TDL01,
Wa05,HMMOV06,HMT06,OH06,KTYI07,
C07,OH08,IHHH08,MW08,DFXY09,H09,JRZZG09a,CVMB10,JRWZGF10,Ba10,N10,KKB11,LW12,CH13a,CLMO13,FLM13,BHLP13}.  
This is because this problem lies at the intersection of two significantly
important topics of quantum information: {\it quantum state
discrimination} \cite{Helstrom76,Holevo82,PW91} and
{\it entanglement theory} \cite{VP07,HHHH09}.  
A quantum state discrimination protocol, which is a protocol for discriminating an
unknown state from known candidates, is an essential subroutine for every quantum
information protocol, since it is the only way to derive classical
information encoded in quantum states.   
On the other hand, entanglement, which is non-local quantum correlation
that does not increase under LOCC, is considered to be an essential resource for
quantum information processing to outperform its classical
counterpart.  
As a result, by studying local discrimination, we can
understand basic quantum information processing among 
spatially separated parties and also the theory of
entanglement itself more deeply.

There are various different settings of state
discrimination problems except the most conventional problem setting 
\cite{Helstrom76,Holevo82,PW91},
like quantum hypothesis testing \cite{HP91,ON00}, quantum
state estimation \cite{Helstrom76,Holevo82}, and classical capacity of
quantum channel \cite{SW96,Holevo98}. 
Thus, there exist various different settings on
local discrimination problems as well. 
In this paper, we especially treat local discrimination problems in the
form of asymptotic hypothesis testing \cite{H06}.

In asymptotic hypothesis testing,
by measuring many copies of an unknown state, 
we aim to certify that the unknown state satisfies a given hypothesis
$H_1$ (called an ``alternative hypothesis''), and for this purpose,
we try to reject hypothesis $H_0$ (called a ``null hypothesis'')
which is true when $H_1$ is false. 
When a hypothesis  ($H_0$ or $H_1$) consists of a single known state, it
is called a simple hypothesis.  In this paper, we only treat asymptotic
hypothesis testing problems whose both null and alternative hypotheses are simple hypotheses.

In asymptotic hypothesis testing, there exist two different error probabilities: 
the error probability judging $H_1$ to be true when $H_0$
is true (type-1 error) and the error probability judging $H_0$ to be 
true when $H_1$ is true (type-2 error). There is a trade-off between these error probabilities, and
hence the way to treat these error probabilities is
not unique. Actually, the following three different optimal error
rates are commonly known and
play important roles in many fields of information theory and
statistical inference \cite{CT91,HK02}: 
\begin{enumerate}
\item  An asymptotic exponent of the optimal type-$2$ error under the
	condition that type-$1$ error is upper bounded by a
	constant. 

\item  An asymptotic exponent of the optimal type-$2$ error under the
	condition that the exponent of the type-$1$ error is bounded by a
	constant. 

\item  An optimum exponent of the average of type-$1$ and type-$2$ error. 
\end{enumerate}
The first optimal error exponent is equal to the relative entropy (or
the Kullback-Leibler divergence) \cite{KL51} between
two hypotheses; this is called ``{\it Stein's lemma}'' \cite{C52}. The
analytical formulas for the second and the third optimal error exponents are
called 
the {\it Hoeffding bound} \cite{H65,B74,CL71} and  {\it Chernoff bound} \cite{C52}, respectively. 
For all three optimal error exponents, their
formulas have been recently extended to the case of the quantum hypothesis
testing, where both $H_0$ and $H_1$ are
single copies of quantum states (say $\rho$ and $\sigma$): the {\it quantum
Stein's lemma} \cite{HP91,ON00},  {\it
quantum Chernoff bound} \cite{ACMBMAV07,NS09}, and  {\it quantum Hoeffding
bound} \cite{H07,N06}.

On the other hand, it is much more difficult to treat an asymptotic quantum hypothesis
testing with an additional locality restriction of a POVM (hereafter
called ``{\it local asymptotic hypothesis testing}''). 
So far, a few papers have treated this problem: the paper by Matthews et al. \cite{MW08} treated the
Chernoff bound under various local POVMs in the case where $\rho$ and $\sigma$ are
completely symmetric and anti-symmetric Werner states, respectively, and
the paper by Nathanson \cite{N10} treated the Chernoff bound under
one-way LOCC POVMs in the case where $\rho$ is a pure bipartite state with a maximal
Schmidt coefficient $\lambda$ and $\sigma$ satisfies
$\Tr \rho \sigma > \lambda$.  After we had published a preprint of 
this paper on arXiv.org \cite{OH11}, two new papers treating asymptotic 
local hypothesis testing appeared.
In order to derive a monogamy type of inequality for the relative 
 entropy of entanglement, Li et al. \cite{LW12} treated 
a hypothesis testing of 
a pair of general tripartite states $\rho_{ABE}$ and $\sigma_{ABE}$
under one-way LOCC POVMs from $A$ to $B$ and studied
how the disturbance on the quantum states induced by the measurement is 
limited when certain type-2 error exponent is achieved.
Most recently, Brandao et al. \cite{BHLP13} calculated
optimal error exponents in the form of Stein's lemma
for a local hypothesis testing between an entangled state
and a set of all separable states.

\subsection{The purpose, results, and organization of this paper}
In the previous paper \cite{OH10}, we treated local hypothesis testing problems
between the completely mixed state $\rho_{mix}$
and an arbitrary fixed bipartite pure state $\ket{\Psi}$ in 
non-asymptotic, or one-copy, settings. 
In particular, we derived analytical formulas of the optimal error probability 
under one-way LOCC and separable operations. 
Although the quantum Chernoff bound \cite{ACMBMAV07,NS09}
can be regarded as the asymptotic version of 
the one-copy formula by Helstrom and Holevo \cite{Helstrom76},
the former requires additional techniques.
That is, the former was obtained from the trace inequality obtained by 
Audenaert et al. \cite{ACMBMAV07,NS09} and a remarkable relation between 
quantum and  classical hypothesis testing by Nussbaum et al. \cite{NS09}. 
The main purpose of this paper is
to invent additional techniques
and to derive the asymptotic bounds for local hypothesis testing based 
on the previous paper \cite{OH10}.
Hence, in this paper, we treat the all three optimal error exponents 
(the Stein, Chernoff, Hoeffding types) 
of asymptotic local hypothesis testing between
the completely mixed state $\rho=\rho_{mix}$
and an arbitrary fixed bipartite pure state $\sigma=\ket{\Psi}\bra{\Psi}$.

As a class of POVMs, we treat one-way LOCC (local operations and one-way classical
communication) POVMs, two-way LOCC (local operations and two-way classical
communication) POVMs and also separable POVMs (POVMs 
implementable by separable operations).   
There may be a controversy as to whether a separable operation is ``{\it local}'', 
since entanglement is necessary to implement non-LOCC separable 
operations \cite{BDFMRSSW99}.  
On the other hand, separable operations cannot generate any entanglement 
from a separable state. Moreover, one of the author recently showed that 
all separable operations can be implemented by local quantum operations 
with  classical correlation which does not have global causal structure 
\cite{AOKM14}. Thus, separable operations surely have a local nature. 
Furthermore, in technical viewpoint, as we will describe in this paper, separable operations often 
approximate two-way LOCC very well when we focus on a particular type of 
information processing.
Hence, we treat separable operations in this paper.
We further treat exponents of global operations, which are already known 
\cite{HP91,ACMBMAV07,H07}.
Comparing the global case and the respective local cases,
we can clarify the effects of the respective local restrictions.

Here, we list the main results of this paper derive the following results, 
where $d_A$ and $d_B$ are local dimensions of a bipartite Hilbert 
space. This list explains how locality affects these kinds of measurements.: 
\begin{enumerate}
\item  The Stein's lemma type of error exponents are
the same for all three classes of local POVMs and given as $\log d_A +
\log d_B - E(\ket{\Psi})$, where $E(\ket{\Psi})$ is the entropy of
       entanglement \cite{VP07,HHHH09}.  
Moreover, their strong converse bound also
       coincides with the optimal error exponents themselves.

\item The Chernoff bound under one-way LOCC POVMs is given as  $\log d_A +
\log d_B - \log R_s(\ket{\Psi})$, 
where $R_s(\ket{\Psi})$ is the Schmidt rank \cite{VP07,HHHH09}. The Chernoff bound under separable POVMs is given as $\log
      d_A + \log d_B - LR(\ket{\Psi})$, where $LR(\ket{\Psi})$ is
      the logarithmic robustness of entanglement \cite{Brandao05,Datta09}.

\item An analytical formula of the Hoeffding bounds  is
      derived under one-way LOCC POVMs without any restriction on a parameter 
      and under separable POVMs for a restricted parameter region.  
For other parameter regions, analytical upper bounds and lower bounds of Hoeffding bounds under
      separable POVMs are derived.   

\item The Chernoff and the Hoeffding bounds under a class of three-step 
       (therefore, two-way) LOCC protocols are numerically calculated 
       for  low-dimensional
      systems. As as result, we show that these bounds not only
      outperform the bounds for one-way LOCC POVMs but also almost approximate the
      bounds for separable POVMs in the parameter region where
      analytical bounds for separable POVMs are derived. 
\end{enumerate}
Note that the base of all logarithms ($\log$) in this paper is 
$e$; that is, it is a natural logarithm. 
Among the above results, result $2$ is remarkable since it  gives a new 
operational meaning for the logarithmic 
robustness of entanglement in terms of this local asymptotic hypothesis
testing problem; for another operational meaning of logarithmic
robustness, see \cite{Brandao07}.  Result $4$ is also remarkable since, as far as we
know, this is first time a gap between optimal error exponents has been 
found under one-way LOCC and two-way LOCC POVMs in asymptotic local discrimination problems;
that is, so far, all such gaps have been bound find in optimal error probabilities in
non-asymptotic local discrimination problems \cite{GV01,C07}, and it was not known whether
such gaps survive in their asymptotic extensions.

This paper is organized as follows: In section \ref{sec
preliminary}, we present mathematical descriptions of our hypothesis
testing problem,  known results  about optimal error exponents under
global POVM, and a short description of main results. Then, we treat the hypothesis
testing problem under one-way LOCC and separable operations in sections
\ref{sec one-way} and \ref{sec separable},  and give analytical
expressions of optimal error exponents under these classes of POVM. 
In section \ref{sec two-way}, we analyze a special class of three-step
LOCC (thus, two-way LOCC) protocols for this local hypothesis testing
problem. In section \ref{section plot}, we present and discuss plots of error exponents 
corresponding to the Chernoff and Hoeffding bounds in low-dimensional systems. 
Finally, we summarize the results of our paper in section
\ref{sec summary}.
We also provide a list of all notations used in this paper in appendix 
\ref{sec notations} for convenience.

\section{Preliminary and main results} \label{sec preliminary}
\subsection{Preliminary}
In this paper, we treat a bipartite quantum system and its
$n$-copy extension. A single copy of a bipartite Hilbert space is
written as 
$\Hi_{AB} \stackrel{\rm def}{=} \Hi _A \otimes \Hi_B$, 
and its local dimensions are written as  $d_A \stackrel{\rm def}{=} \dim \Hi_A$ 
and $d_B \stackrel{\rm def}{=} \dim \Hi_B$. A space of all operators on 
a Hilbert space $\Hi$ is written as $\B \left(\Hi\right)$. We use 
notations such as 
$I_A$, $I_B$, $I_{AB}$, $I_A^n$, $I_B^n$, and $I_{AB}^n$ for identity
operations on $\Hi_A$, $\Hi_B$, $\Hi _{AB}$, $\Hi_A^{\otimes n}$,
$\Hi_B^{\otimes n}$, and $\Hi_{AB}^{\otimes n}$, respectively. 
When it is  easy to identify the support of an identity operator, we
abbreviate them as $I$ hereafter. 

In this paper, we consider an asymptotic hypothesis testing between
$n$-copies of an arbitrary fixed pure-bipartite
state $\ket{\Psi}$ having Schmidt decomposition as 
\begin{equation}\label{eq schmidt decomposisiton psi}
\ket{\Psi}\stackrel{\rm def}{=}\sum_{i=1}^{d}
\sqrt{\lambda_i}\ket{i}\otimes\ket{i},
        \end{equation}
 where $d$ is defined by 
\begin{equation}\label{eq def d}
			  d
			 \stackrel{\rm def}{=} \min (d_A,d_B),
\end{equation}
and $n$-copies of the completely
mixed state (or a white noise) by
\begin{equation}\label{eq def rho mix}
\rho _{mix}\stackrel{\rm def}{=}\frac{I_{AB}}{d_Ad_B}
\end{equation} 
under the various 
restrictions on available POVMs: global POVMs, separable POVMs, one-way
LOCC POVMs, two-way LOCC POVMs \cite{VP07,HHHH09}. 
We choose the completely mixed state
$\rho_{mix}^{\otimes n}$ as
a null hypothesis and the state $\ket{\Psi}^{\otimes n}$ as an
alternative hypothesis. 
In the following part of this section, we give definitions of various error exponents and optimal
error exponents for a general simple null hypothesis $\rho^{\otimes n}$  and
an alternative
hypothesis $\sigma^{\otimes n}$.
Thus, $\rho=\rho_{mix}$ and $\sigma=\Psi\stackrel{\rm def}{=}\ket{\Psi}\bra{\Psi}$ in our local
hypothesis testing problem.

We only treat a two-valued POVM consisting of POVM elements $T_n$
and $I^n_{AB}-T_n$, where $T_n$ is supported by 
$\Hi_A^{\otimes n} \otimes \Hi_B^{\otimes n}$.
When the measurement result is $T_n$, we judge an unknown state as
$\sigma^{\otimes n}$,
and when the measurement result is $I_{AB}^n-T_n$, we judge the unknown state
as $\rho^{\otimes n}$.

Thus, type-1 error is written as 
\begin{equation}\label{eq def type 1 error}
\alpha _n(T_n)= \Tr \rho^{\otimes n}T_n,
\end{equation}
and type-2 error is written as 
\begin{equation}\label{eq def type 2 error}
\beta _n (T_n) = \Tr \sigma^{\otimes n}  \left ( I_{AB}^n-T_n \right ).
\end{equation}
As a result, the optimal type-2 error under the condition 
that the type-1 error is no more than a constant $\alpha\ge 0$ is written as
\begin{align}\label{eq def beta n C alpha rho sigma}
&\beta_{n,C}(\alpha|\rho \| \sigma) \nonumber \\ &\stackrel{\rm def}{=}  
 \min _{T_n} \left \{
 \beta_n (T_n) \  | \ \alpha_n (T_n) \le \alpha, \{ T_n, I-T_n \} \in C \right \},
\end{align}
where $C$ is either $\rightarrow$, $\leftrightarrow$, $Sep$, or $g$
corresponding to classes of one-way LOCC, two-way LOCC, separable
and global POVMs, respectively.
Here, based on the original definitions of these classes,
we notice that $\leftrightarrow$ is not compact although $\rightarrow$, 
$Sep$, or $g$ are compact sets \cite{CLMOW13}.
Hence, in this paper, 
class $\leftrightarrow$ is defined as a closure of the set of all two-way LOCC 
POVMS, which involves infinite-step LOCC protocols as well \cite{BDFMRSSW99,KKB11,OBNM08,Ch11,CCL12}.
This modified definition of class $\leftrightarrow$ justifies the use of 
$min$ in Eq.~(\ref{eq def beta n C alpha rho sigma}) in the case of 
$C=\leftrightarrow$.  
Similarly, the optimal type-1 error under the condition 
that the type-2 error is no more than a constant $\beta \ge 0$ is written as
\begin{align}\label{eq def alpha n C alpha rho sigma}
&\alpha_{n,C}(\beta|\rho\| \sigma) \nonumber \\& \stackrel{\rm def}{=}  \min _{T_n} \left \{
 \alpha_n (T_n) \  | \ \beta_n (T_n) \le \beta, \{ T_n, I-T_n \} \in C \right \}.
\end{align}
It is easily shown that the function $\beta \mapsto \alpha_{n,C}(\beta|\rho\| \sigma)$ is an inverse function of
the function $\alpha \mapsto \beta_{n,C}(\alpha|\rho\| \sigma)$ in the region where $\beta_{n,C}(\alpha|\rho\| \sigma)$ is
strictly decreasing and continuous \cite{OH10}.

In the Bayesian problem setting, we further assume 
the existence of a prior probability on hypotheses. 
Suppose there exists a prior probability $(\pi_0, \pi_1)$ on the null and
alternative hypotheses. Then, the mean error probability is given as $\pi_0 \alpha_n (T_n) + \pi_1 \beta_n (T_n)$. 
Thus, for a given class of POVMs $C$, the optimal mean error probability is defined as 
\begin{align}\label{eq def p n c pi 0 pi 1 rho sigma}
& P_{n,C}(\pi_0,\pi_1|\rho \| \sigma) \nonumber \\
&\stackrel{\rm def}{=} \min_{T_n} \left \{
  \pi_0 \alpha_n (T_n) + \pi_1 \beta_n (T_n)| \{ T_n, I-T_n \} \in C \right \}.
\end{align}
For any class
of POVMs $C$,  the relation between $P_{n,C}(\pi_0,\pi_1|\rho \| \sigma)$ and $\beta_{n,C}(\alpha|\rho \| \sigma)$ is given as follows:
\begin{equation}\label{eq P_{n,C} = inf alpha beta}
 P_{n,C}(\pi_0,\pi_1|\rho \| \sigma)=\inf_{0 \le \alpha \le 1} \pi_0 \alpha + \pi_1 \beta_{n,C}(\alpha|\rho\| \sigma).
\end{equation}
Similarly, the following formula holds between $P_{n,C}(\pi_0,\pi_1|\rho \| \sigma)$ and $\alpha_{n,C}(\beta|\rho \| \sigma)$:
\begin{equation}\label{eq P_{n,C} = inf alpha beta 2}
 P_{n,C}(\pi_0,\pi_1|\rho \| \sigma)=\inf_{0 \le \beta \le 1} \pi_0 \alpha_{n,C}(\beta|\rho \| \sigma) + \pi_1 \beta.
\end{equation}
Eqs.~(\ref{eq P_{n,C} = inf alpha beta}) and (\ref{eq P_{n,C} = inf 
alpha beta 2}) hold even when a null hypothesis and an alternative
hypothesis are composite hypotheses.

In this paper, we treat this local hypothesis testing problem in the
forms of the Chernoff bound, the Stein's Lemma, and the Hoeffding bound.
For a class of POVMs $C$, the Chernoff bound $\xi_{C}\left(\rho \|\sigma 
\right)$ is given as the optimal exponent of the mean error
probability: 
\begin{equation}\label{eq def xi C}
 \xi_{C}(\rho\| \sigma) \stackrel{\rm def}{=} \lim_{n \rightarrow \infty} - \frac{1}{n} \log P_{n,C}(\pi_0,\pi_1|\rho\| \sigma).
\end{equation}
Hence, for a large $n$, $P_{n,C}\left(\pi_0,\pi_1|\rho \|\sigma \right)$ behaves like 
\begin{equation}
 P_{n,C}\left(\pi_0,\pi_1|\rho \|\sigma \right)\sim \exp 
\left(-n\xi_C\left(\rho\| \sigma \right)\right).
\end{equation}
$\xi_{C}(\rho\|\sigma)$ may not exist for a given class $C$. Hence, we 
define $\overline{\xi}_{C}(\rho \|\sigma)$ and $\underline{\xi}_{C}(\rho 
\|\sigma)$ by using $\overline{\lim}$ or $\underline{\lim}$ instead of 
$\lim$, where $\overline{\lim}$ and $\underline{\lim}$ are the 
limit-superior and the limit-inferior, 
respectively. $\overline{\xi}_{C}(\rho \|\sigma)$ 
and $\underline{\xi}_{C}(\rho \|\sigma)$ alway exist and satisfy 
$\overline{\xi}_{C}(\rho \|\sigma) \ge 
\underline{\xi}_{C}(\rho\|\sigma)$. $\xi_{C}(\rho \|\sigma)$ exists if 
and only if  $\overline{\xi}_{C}(\rho \|\sigma) = 
\underline{\xi}_{C}(\rho\|\sigma)$. When the class of global 
POVMs $C=g$, it is known that the Chernoff bound $\xi_{g}(\rho\|\sigma)$ 
is given as \cite{ACMBMAV07}
\begin{equation}\label{eq analytical formula global chernoff}
 \xi_{g}(\rho\| \sigma) = -\log \left( \inf _{0 \le s \le 1} \Tr \left(
						      \rho^{1-s}\sigma^s
						     \right)\right).
\end{equation}

In an asymmetric hypothesis testing, we are interested
in the optimal type-$2$ (or $1$) error exponent under various restrictions
for the type-$1$ (or $2$) error. 
If the error exponent exists, we define the optimal type-$2$ (or $1$) 
error exponent $\theta_{C}\left(\epsilon|\rho \|\sigma\right)$
(or $\eta_{C}\left(\epsilon|\rho \|\sigma\right)$)
under the restriction where the
optimal type $2$ (or $1$) is less than a constant $\epsilon$ as
\begin{align}
 \theta_{C}(\epsilon|\rho\| \sigma) \stackrel{\rm def}{=}& \lim_{n \rightarrow \infty}
 -\frac{1}{n}\log \beta_{n,C}(\epsilon|\rho\| \sigma), \label{eq def stein beta}\\ 
\eta_{C}(\epsilon|\rho\| \sigma) \stackrel{\rm def}{=}& \lim_{n \rightarrow \infty}
 -\frac{1}{n}\log \alpha_{n,C}(\epsilon|\rho\| \sigma).\label{eq def stein alpha}
\end{align} 
Hence, for a large $n$, $\beta_{n,C}\left(\epsilon|\rho \|\sigma 
\right)$ and $\alpha_{n,C}\left(\epsilon|\rho \|\sigma \right)$ 
behave like 
\begin{align}
 \beta_{n,C}\left(\epsilon|\rho \|\sigma 
\right)\sim \exp 
\left(-n\theta_C\left(\epsilon|\rho\| \sigma \right)\right)
\end{align}
and
\begin{align}
 \alpha_{n,C}\left(\epsilon|\rho \|\sigma 
\right)\sim \exp 
\left(-n\eta_C\left(\epsilon|\rho\| \sigma \right)\right),
\end{align}
respectively.
We define $\overline{\theta}_{C}(\epsilon|\rho\| \sigma)$, 
$\underline{\theta}_{C}(\epsilon|\rho\| \sigma)$, 
$\overline{\eta}_{C}(\epsilon|\rho\| \sigma)$, and 
$\underline{\eta}_{C}(\epsilon|\rho\| \sigma)$ by using
$\overline{\lim}$ and  $\underline{\lim}$, respectively, instead of
$\lim$. 
By the definitions, $\theta_{C}(\epsilon|\rho\| \sigma)$ exists if and only if 
$\overline{\theta}_{C}(\epsilon|\rho\| \sigma)=\underline{\theta}_{C}(\epsilon|\rho\| \sigma)$.
Similarly, $\eta_{C}(\epsilon|\rho\| \sigma)$ exists if and only if 
$\overline{\eta}_{C}(\epsilon|\rho\| \sigma)=\underline{\eta}_{C}(\epsilon|\rho\| \sigma)$. 

For a class of POVMs $C$, we define the
strong converse bounds of $\theta_{C}(\epsilon|\rho \|\sigma)$ 
as  
\begin{align}
\quad& \theta_{C}^\dagger(\rho \|\sigma)
 \nonumber \\ 
\stackrel{\rm def}{=}& \sup_{\{T_n\}_{n = 1}^{\infty}} \Big \{ \underline{\lim}_{n \rightarrow
 \infty } -\frac{1}{n} \log \beta_{n}(T_n) \ \Big |
 \nonumber \\
\quad& \qquad \underline{\lim}_{n \rightarrow \infty} \alpha_{n}(T_n)
 < 1, \{T_n, I -T_n\}
 \in C \Big \}. \label{eq def strong converse}
\end{align}
This value is the bound of the optimal type-1 error exponent of a
sequence of POVM $\{ T_n \}$ whose type-2 error probability 
does not converge to $1$. 
$\eta_{C}^\dagger(\rho \|\sigma)$ is also defined as
\begin{align}
\quad& \eta_{C}^\dagger(\rho \|\sigma)
 \nonumber \\ 
\stackrel{\rm def}{=}& \sup_{\{T_n\}_{n = 1}^{\infty}} \Big \{ \underline{\lim}_{n \rightarrow
 \infty } -\frac{1}{n} \log \alpha_{n}(T_n) \ \Big |
 \nonumber \\
\quad& \qquad \underline{\lim}_{n \rightarrow \infty} \beta_{n}(T_n)
 < 1, \{T_n, I -T_n\}
 \in C \Big \}. \label{eq def strong converse eta}
\end{align} 
By the definition, when $\eta_{C}(\epsilon|\rho \|\sigma)$ and
$\theta_{C}(\epsilon|\rho \|\sigma)$   exist, we derive 
\begin{align*}
 \eta_{C}(\epsilon|\rho \|\sigma) &\le \eta_{C}^\dagger(\rho \|\sigma), \\
 \theta_{C}(\epsilon|\rho \|\sigma) &\le \theta_{C}^\dagger(\rho \|\sigma)
\end{align*}
for all $\epsilon>0$ and any class $C$.

It is known that $\theta_{g}(\epsilon|\rho \|\sigma)$ and 
$\theta_{g}^\dagger (\rho \|\sigma)$ are
given for a null $\rho^{\otimes n}$ and an alternative
$\sigma^{\otimes n}$ hypotheses as \cite{HP91,ON00}
\begin{align}\label{eq global stein}
 &\theta_{g}(\epsilon|\rho \|\sigma)=\theta_{g}^\dagger (\rho \|\sigma)=
 \eta_{g}(\epsilon|\sigma\| \rho)= \eta_g^{\dagger}(\sigma\|
 \rho) \nonumber \\
= &D(\rho \| \sigma ),
\end{align}
where $D(\rho \|\sigma)$ is the relative entropy between $\rho$ and
$\sigma$ defined as \cite{U62}
\begin{equation}\label{eq def relative entropy}
 D\left(\rho \|\sigma \right)\stackrel{\rm def}{=}\Tr \rho \log \rho 
  -\Tr \rho \log \sigma.
\end{equation}
This result is called Stein's lemma.

For a class of POVMs $C$, the Hoeffding bound $B_C(r|\rho\|\sigma)$ (or $A_C(r|\rho\|\sigma)$) is
defined as the optimum type-$2$ (or type-$1$) error exponent under the
restriction where the other error exponent is lower bounded by a constant $r$:
\begin{align}
 B_C(r|\rho\|\sigma) &\stackrel{\rm def}{=} \sup _{\left\{ T_n 
 \right\}_{n=1}^\infty}  \{ \lim _{n \rightarrow
 \infty} -\frac{1}{n}\log \beta_{n}(T_n) \big |  \nonumber \\
&  \lim _{n \rightarrow
 \infty} -\frac{1}{n}\log \alpha_{n}(T_n) \ge r, \{T_n,I-T_n\}\in C  \}, 
 \label{eq def b c}
\\
 A_C(r|\rho\|\sigma) &\stackrel{\rm def}{=} \sup _{\left\{ T_n 
 \right\}_{n=1}^\infty} \{ \lim _{n \rightarrow
 \infty} -\frac{1}{n}\log \alpha_{n}(T_n) \big |  \nonumber \\
& \lim _{n \rightarrow
 \infty} -\frac{1}{n}\log \beta_{n}(T_n) \ge r, \{T_n,I-T_n\}\in C  \}. \label{eq def a c}
\end{align}
In the above equations, type-1 error probability 
$\alpha\left(T_n\right)$ and type-2 error probability 
$\beta\left(T_n\right)$ are defined by Eqs.~(\ref{eq def type 1 error}) 
and (\ref{eq def type 2 error}), respectively.
For global POVMs $C=g$,  the Hoeffding bounds have the following formula \cite{H07,ANSV08}:
\begin{equation}\label{eq global hoeffding}
B_g(r|\rho\|\sigma )= A_g(r|\sigma \|\rho)=\sup _{0\le s<1} \frac{-rs-\log \Tr \sigma^s \rho^{1-s}}{1-s}. 
\end{equation}

By means of the above analytical formulas for the various optimal error exponents
under global POVM, we can calculate these optimal global error exponents for
our hypothesis testing problem as follows: 
\begin{align}
& \xi_g(\rho_{mix}\|\Psi) =
 \eta_{g}(\epsilon|\rho_{mix}\|\Psi)=A_g(r|\rho_{mix}\|\Psi) \nonumber \\
&\qquad \qquad \quad  =\log d_A + \log d_B, \ (\forall \epsilon>0, \forall r 
 >0)\\
& \theta_{g}(\epsilon|\rho_{mix}\|\Psi) = +\infty, \ (\forall \epsilon>0)\label{eq analytical
 global stein beta}\\
& B_g(r|\rho_{mix}\|\Psi) \nonumber \\
&=\left \{
\begin{array}{l}
+\infty  \quad {\rm if} \quad 0 \le r \le \log d_A + \log d_B, 
\\
\quad  \\
0 \quad {\rm otherwise}. \\
\end{array}
\right .
\end{align}
From the definition of 
$\theta_{g}(\epsilon|\rho_{mix}\|\Psi)$ in Eq.~(\ref{eq def stein beta}), Eq.~(\ref{eq analytical 
global stein beta}) shows the super-exponential convergence
of the optimal type-2 error $\beta_{n,g}(\epsilon|\rho_{mix}\|\Psi)$. 
Moreover, one can easily see that for a fixed $\alpha$, 
for $n \ge -\frac{\log \epsilon}{\log d_A + \log d_B}$, the optimal type-2 error
$\beta_{n,g}(\epsilon|\rho_{mix}\|\Psi)$ is exactly equal to $0$.

Finally, we slightly explain the optimal tests of
classical hypothesis testings. As we can easily see,  
when $\rho$ and $\sigma$ are diagonal in a fixed basis 
$\{\ket{i}\}_{i=1}^d$, Eqs.(\ref{eq analytical formula global chernoff}), 
(\ref{eq global stein})
, and (\ref{eq global hoeffding})
give the optimal error exponents of the Stein, Chernoff, and Hoeffding 
types of classical hypothesis testings, respectively.
The optimal classical measurements achieving these bounds can be chosen as 
``{\it likelihood-ratio tests}'' $\Lambda\left( t \right)$ \cite{CT91,B74}. 
Let us consider a hypothesis testing of $n$-copy-independent and identical distributions
$p\left(x_n\right)$ and $q\left(x_n\right)$, where $x_n$  is on a set 
$\chi_n$ of all sequences of $n$-alphabets. Then, the likelihood-ratio 
test $\Lambda \left(t \right)$ is defined by 
\begin{equation}\label{eq def lrt}
 \Lambda \left( t \right)\stackrel{\rm def}{=} \left\{x_n \in 
	  \chi_n \big |\frac{p\left( x_n \right)}{q\left(x_n\right) }>t \right\}.
\end{equation}
Hence,  if the measurement result $x_n$ is in $\Lambda \left(t \right)$, then, 
we judge an unknown distribution as $p\left(x_n\right)$, and otherwise, 
we judge it as $q\left(x_n\right)$.
The optimal tests whose error exponents converge to the Stein, Chernoff and Hoeffding type 
exponents are known to be $\Lambda \left(t \right)$ with
$t=2^{n\left( D\left(p\left(x_n \right)| q \left(x_n 
\right)\right)-\delta\right)}$, $1$, and 
$2^{n\left(B_g\left(r|p\left(x_n\right)\|q\left(x_n\right)\right)-r\right)}$, respectively,
where $\delta>0$ is an arbitrary small parameter \cite{CT91,B74}.

\subsection{Main results}
In this subsection, we briefly describe the main 
results of this paper. 
The Stein's lemma type of error exponents as well as their 
strong converse bounds are given for all $0 <\epsilon < 1$ as 
(Theorem \ref{theorem one way}, Theorem \ref{theorem sec asymptotic}, 
and Corollary \ref{corollary strong converse}),
\begin{align}
&\theta_{\rightarrow}\left(\epsilon|\rho_{mix}\|\Psi \right) = 
 \theta_{\leftrightarrow}\left(\epsilon|\rho_{mix}\|\Psi \right)  = 
 \theta_{sep}\left(\epsilon|\rho_{mix}\|\Psi \right)  \nonumber 
 \\
=&\theta^{\dagger}_{\rightarrow}\left(\rho_{mix}\|\Psi \right) 
=\theta^{\dagger}_{\leftrightarrow}\left(\rho_{mix}\|\Psi \right)
=\theta^{\dagger}_{sep}\left(\rho_{mix}\|\Psi \right)
\nonumber 
\\=& +\infty, \\
&\eta_{\rightarrow}\left(\epsilon|\rho_{mix}\|\Psi \right) = 
 \eta_{\leftrightarrow}\left(\epsilon|\rho_{mix}\|\Psi \right)  = 
 \eta_{sep}\left(\epsilon|\rho_{mix}\|\Psi \right)  \nonumber 
 \\
=&\eta^{\dagger}_{\rightarrow}\left(\rho_{mix}\|\Psi \right) 
=\eta^{\dagger}_{\leftrightarrow}\left(\rho_{mix}\|\Psi \right)
=\eta^{\dagger}_{sep}\left(\rho_{mix}\|\Psi \right)
\nonumber 
\\= &\log d_A + \log d_B -E\left(\ket{\Psi} \right),
\end{align}
where $E\left(\ket{\Psi}\right)$ is the entropy of the entanglement 
defined by Eq.~(\ref{eq def entropy of entanglement}).
Note that the difference between the class of POVMs does not appear from 
the viewpoint of the Stein's lemma type of error exponents. 

The Chernoff bounds for one-way LOCC POVMs and separable POVMs are given 
as (Theorem \ref{theorem one way} and Theorem \ref{theorem xi _sep le chernoff of simple})
\begin{align}
 \xi_{\rightarrow}(\rho_{mix}\|\Psi) & = \log d_A +\log d_B -\log 
  R_{s}(\ket{\Psi})\nonumber \\
 \xi_{sep}(\rho_{mix} \|\Psi) 
&= \log d_A + \log d_B -LR(\ket{\Psi}), 
\end{align}
where $R_{s}(\ket{\Psi})$ is the Schmidt rank defined by Eq.~(\ref{eq def schmidt rank}),
and $LR(\ket{\Psi})$ is the logarithmic robustness of entanglement defined 
by Eq.~(\ref{eq def LR}). 
The above single-letterized formulas show the clear separation between
the Chernoff bounds under one-way LOCC and under separable operations,
which never appears for the  Stein's lemma type of error exponents.
For two-way LOCC POVMs $C=\leftrightarrow$, in 
Section \ref{sec two-way}, we 
derive the bound $\tilde{\xi}_{\leftrightarrow} \left(\rho_{mix} \| \Psi 
\right)$, which can be numerically calculated by Eq.~(\ref{eq tilde xi = sup s} ),
 satisfying
 \begin{align*}
\xi_{\rightarrow}  \left(\rho_{mix} \| \Psi \right) \le & 
 \tilde{\xi}_{\leftrightarrow} \left(\rho_{mix} \| \Psi \right)  \le
  \xi_{\leftrightarrow}\left(\rho_{mix} \| \Psi \right). 
 \end{align*} 
In Section \ref{section plot}, we will show that,  
although $\tilde{\xi}_{\leftrightarrow}\left(\rho_{mix}\|\Psi \right)$
is the Chernoff bound of the restricted class of three-step LOCC 
protocols (Alice$\rightarrow$Bob$\rightarrow$Alice), 
it not only 
exceeds $\xi_{\rightarrow}  \left(\rho_{mix} \| \Psi \right)$ but also 
well
 approximates $\xi_{sep}\left(\rho_{mix}\|\Psi \right)$ at least in  
 low-dimensional Hilbert spaces (FIG.~1 and 2). 

The Hoeffding bounds for one-way LOCC POVMs are given as
follows (Theorem \ref{theorem one way}):
\begin{align*}
 & A_{\rightarrow}(r|\rho_{mix}\|\Psi) \nonumber \\
&  =\log d_A+\log d_B + \sup _{0\le s <1} \frac{-rs-\log \sum_i \lambda_i^s}{1-s},\\
& B_{\rightarrow}(r|\rho_{mix}\|\Psi) \nonumber \\
&=\left \{
\begin{array}{l}
+\infty  \quad {\rm if} \quad 0 \le r \le \log d_A + \log d_B + \log R_s,
\\
\quad  \\
{\rm otherwise} \\
\sup _{0 \le s <1} \frac{-(r-\log d_A -\log d_B)s - \log \sum
 _{i}\lambda_i^{1-s}}{1-s}.
\end{array}
\right .
\end{align*}
The Hoeffding bounds for separable POVM are evaluated as (Theorem 
\ref{theorem capital b_sep r} and Corollary \ref{corollary A sep r})
\begin{align*}
& A_{sep}(r|\rho_{mix}\|\Psi)= \log d_A + \log d_B - LR(\ket{\Psi}), 
 \nonumber \\
&\qquad \mbox{if } r \ge \log d - LR(\ket{\Psi}), \nonumber \\
&\quad \log d_A +\log d_B - LR(\ket{\Psi}) \le A_{sep}(r|\rho_{mix}\|\Psi)\nonumber \\
& \qquad \qquad \qquad \qquad\le \log
  d_A + \log d_B -E(\ket{\psi}), \nonumber \\
& \qquad \mbox{if }0 \le r \le \log d - LR(\ket{\Psi}), \nonumber \\
&B_{sep}(r|\rho_{mix}\|\Psi)  \nonumber \\
&=\left \{
\begin{array}{l}
+\infty,  \quad {\rm if} \ 0 \le r \le \log d_A + \log d_B - LR(\ket{\Psi}),
\\
\quad  \\
 0, \quad  {\rm if} \ r \ge  \log d_A + \log
 d_B -E(\ket{\Psi}),\\
\end{array}
\right . \\
& B_{sep}(r|\rho_{mix}\|\Psi) \le \log d- LR(\ket{\Psi}), \nonumber \\
& \qquad \mbox{\rm if} \ r >
\log d_A + \log d_B - LR(\ket{\Psi}).
\end{align*}
For two-way LOCC POVM $C=\leftrightarrow$, in 
Section \ref{sec two-way}, we 
derive the bound $\tilde{A}_{\leftrightarrow}\left(r|\rho_{mix} \| \Psi 
\right) $, which can be numerically calculated by Eq.~(\ref{eq tilde A = sup s} ),
 satisfying
\begin{align*}
A_{\rightarrow}\left(r|\rho_{mix} \| \Psi \right) \le & \tilde{A}_{\leftrightarrow}\left(r|\rho_{mix} \| \Psi \right) \le
 A_{\leftrightarrow}\left(r|\rho_{mix} \| \Psi \right).
 \end{align*}
In Section \ref{section plot}, we will show that,  
although $\tilde{A}_{\leftrightarrow}\left(r|\rho_{mix} \| \Psi 
\right) $
is the Hoeffding bound of the restricted class of three-step LOCC protocols, 
it  not only 
exceeds $A_{\rightarrow}\left(r|\rho_{mix} \| \Psi \right)$ but also well
 approximates $A_{sep}\left(r|\rho_{mix} \| \Psi \right)$ in the region 
 of parameter $r$ where we have an analytical formula of 
 $A_{sep}\left(r|\rho_{mix} \| \Psi \right)$, at least in the 
 low-dimensional Hilbert spaces (FIG.~3 and 4).

\section{Hypothesis testing under one-way LOCC POVMs} \label{sec one-way}
In this section, we consider  $C=\rightarrow$, that is, the local
hypothesis testing under one-way LOCC POVMs.
In the following part of the paper, we mostly treat the case
where a null hypothesis is $\rho=\rho_{mix}$ and an alternative
hypothesis is $\sigma=\Psi \stackrel{\rm def}{=}\ket{\Psi}\bra{\Psi}$. Therefore, when this is the
case, we often abbreviate these variables in the formula. For example, we write
$\theta_{C}(\epsilon|\rho_{mix}\|\Psi)$
as $\theta_{C}(\epsilon)$.

First, from the last paper \cite{OH10}, we derive the following lemma:
\begin{Lemma}\label{sec one-way lemma}
For all $\alpha >0$ and $\beta >0$,
\begin{align}
\alpha_{n,\rightarrow}
 (\beta|\rho_{mix}\|\Psi)&=\alpha_{n,g}(\beta|\rho_{mix}\|\sigma_{\Psi}), \label{eq
 one-way equal classical 1}\\
\beta_{n,\rightarrow}
 (\alpha|\rho_{mix}\|\Psi)&=\beta_{n,g}(\alpha|\rho_{mix}\|\sigma_{\Psi}), \label{eq
 one-way equal classical 2}
\end{align}
where an optimal type-1 error probability $\alpha_{n,\rightarrow}
 (\beta|\rho \|\sigma)$ and an optimal type-2 error probability $\beta_{n,\rightarrow}
 (\alpha|\rho \|\sigma)$ are defined by Eqs.~(\ref{eq def alpha n C alpha 
 rho sigma}) and   (\ref{eq def beta n C alpha rho sigma}). 
In the above equations,  $\sigma_{\Psi}$ is a separable state defined as 
\begin{equation}\label{eq def sigma psi}
 \sigma_{\Psi}\stackrel{\rm def}{=} \sum_{i=1}^d\lambda_i\ket{i}\bra{i}\otimes \ket{i}\bra{i},
\end{equation}
and $\left\{\ket{i}\otimes \ket{j}\right\}_{i,j}$ is the Schmidt basis 
 of $\ket{\Psi}$ (see Eq.~(\ref{eq schmidt decomposisiton psi})).
\end{Lemma}
({\bf Proof})\\
From Theorem 1 of the last paper \cite{OH10}, 
an optimal one-way LOCC POVM achieving $\beta_{1,\rightarrow}(\alpha|\rho_{mix}\|\Psi)$
can be chosen to be
 diagonal in the Schmidt basis of $\ket{\Psi}$ 
(thus, $\{ \ket{i}\otimes \ket{j} \}_{ij}$ in this case).
This is also true for $\beta_{n,\rightarrow}(\alpha|\rho_{mix}\|\Psi)$.
Thus, in this case, an optimal one-way LOCC POVM element $T_n$ can be chosen to be diagonal in
the Schmidt basis of $\ket{\Psi}^{\otimes n}$; 
we use the notation $\{\ket{J_n} \}_{J_n=1}^{d^n}$ as an abbreviation of the
basis $\{ \ket{i_1} \otimes \cdots \otimes \ket{i_n}\}_{i_1,\dots,i_n}$.
  
All the Schmidt-basis-diagonal POVM elements $T_n$ are  one-way LOCC
POVM elements, and can be written  in the form 
\begin{align}
T_n=  \sum _{J_n,J'_n=1}^{d^n} 
&\ket{J_n}\bra{J_n} \otimes
 \ket{J'_n}\bra{J'_n}
T'_n \nonumber \\
&\ket{J_n}\bra{J_n} \otimes
 \ket{J'_n}\bra{J'_n},
\end{align}
where $T'_n$ is a global POVM element.
On the other hand, for an arbitrary global POVM element $T'_n$, 
the above $T_n$ is a one-way LOCC POVM element.
For such a POVM element $T_n$, the error probabilities 
$\alpha_n(T_n)$ and $\beta_n(T_n)$, defined by Eqs.~(\ref{eq 
def type 1 error}) and (\ref{eq def type 2 error}), respectively, can be written as 						      
 \begin{align}
\alpha_n(T_n) &= \Tr \rho_{mix}^{\otimes n}T'_n, \\
  \beta_n(T_n)&=\Tr \sigma_{\Psi}^{\otimes n}T'_n.
 \end{align}
Substituting the above equations into Eqs.~(\ref{eq def beta n C alpha rho sigma})
and (\ref{eq def alpha n C alpha rho sigma}), we derive 
Eqs.~(\ref{eq
 one-way equal classical 1}) and (\ref{eq
 one-way equal classical 2}). \hfill $\square$ \\

Here, note that since $\rho_{mix}$ and $\sigma_\Psi$ are commutative, 
the quantum global hypothesis testing between $\rho_{mix}$ and $\sigma_\Psi$ are equivalent 
to the classical hypothesis testing between the probability distributions
$p_{mix}(i,j)$ and $q_\Psi(i,j)$, where $p_{mix}(i,j)$ and $q_\Psi(i,j)$
are defined as $p_{mix}(i,j)=\frac{1}{d_Ad_B}$ and 
$q_\Psi(i,j)=\lambda_i \delta_{i,j}$, respectively.
Hence, Lemma \ref{sec one-way lemma} guarantees that the quantum 
hypothesis testing between $\rho_{mix}$ and $\ket{\Psi}$ by one-way LOCC POVMs
can be reduced to the classical hypothesis testing between $p_{mix}(i,j)$ and $q_\Psi(i,j)$.

By means of the above lemma, we derive analytical formulas for 
the optimal error exponents: 
\begin{Theorem}\label{theorem one way}
For all $\epsilon>0$ and $r>0$, 
 \begin{align}
&  \xi_{\rightarrow}(\rho_{mix}\|\Psi)  = \log d_A +\log d_B -\log 
  R_{s}(\ket{\Psi}), \label{eq xi rightarrow}\\
&\eta_{\rightarrow }(\epsilon|\rho_{mix}\|\Psi) = \log d_A+\log d_B
  -E(\ket{\Psi}), \label{eq alpha R epsilon rightarrow}\\
& \theta_{\rightarrow}(\epsilon|\rho_{mix}\|\Psi) = + \infty, \\
& A_{\rightarrow}(r|\rho_{mix}\|\Psi) \nonumber \\
&  =\log d_A+\log d_B \quad + \sup _{0\le s <1} \frac{-rs-\log \sum_i \lambda_i^s}{1-s},\\
& B_{\rightarrow}(r|\rho_{mix}\|\Psi) \nonumber \\
&=\left \{
\begin{array}{l}
+\infty  \quad {\rm if} \quad 0 \le r \le \log d_A + \log d_B + \log R_s,
\\
\quad  \\
{\rm otherwise} \\
\sup _{0 \le s <1} \frac{-(r-\log d_A -\log d_B)s - \log \sum
 _{i}\lambda_i^{1-s}}{1-s}.
\end{array}
\right .
 \end{align}
 In the above formulas, $\xi_{\rightarrow}(\rho\|\sigma)$, 
$\eta_{\rightarrow }(\epsilon|\rho\|\sigma)$, 
$\theta_{\rightarrow}(\epsilon|\rho\|\sigma)$
$A_{\rightarrow}(r|\rho\|\sigma)$, and $B_{\rightarrow}(r|\rho\|\sigma)$ 
 are defined by Eqs.~(\ref{eq def xi C}), (\ref{eq def stein alpha}), (\ref{eq def stein beta}), 
(\ref{eq def a c}), and (\ref{eq def b c}), respectively.
$R_s(\ket{\Psi})$ is the Schmidt rank defined as 
\begin{equation}\label{eq def schmidt rank}
 R_s(\ket{\Psi})\stackrel{\rm def}{=} {\rm rank} \sigma_A,
\end{equation}
where $\sigma_A$ is defined as $\sigma_A\stackrel{\rm def}{=} \Tr_B \ket{\Psi}\bra{\Psi}$, 
and
 $E(\ket{\Psi})$ is the entropy of entanglement of $\ket{\Psi}$ defined 
 as \cite{NC00,H06,VP07,HHHH09}
\begin{equation}\label{eq def entropy of entanglement}
 E\left(\ket{\Psi}\right)\stackrel{\rm def}{=}-\Tr \sigma_A\log \sigma_A.
\end{equation}
\end{Theorem}
{\bf (Proof)} \\
Eqs.~(\ref{eq P_{n,C} = inf alpha beta}) and (\ref{eq
 one-way equal classical 2}) [see Lemma \ref{sec one-way lemma}] guarantee 
\begin{equation}
 P_{n,\rightarrow}(\pi_0,\pi_1|\rho_{mix}\|\Psi)= P_{n,g}(\pi_0,\pi_1|\rho_{mix}\|\sigma_{\Psi}).
\end{equation}
The above equation and the definition of $\xi_{C}(\rho\|\sigma)$ 
(see Eq.~(\ref{eq def xi C})) guarantee
\begin{equation}
 \xi_{\rightarrow}(\rho_{mix}\|\Psi)= \xi_{g}(\rho_{mix}\|\sigma_{\Psi}).
\end{equation}
Similar equations also hold for $\eta_{C}(\epsilon|\rho_{mix}\|\Psi)$, $\theta_{C}(\epsilon|\rho_{mix}\|\Psi)$,
$A_C(r|\rho_{mix}\|\Psi)$ and $B_C(r|\rho_{mix}\|\Psi)$. Since analytical formulas are known for these optimum
error exponents under global POVMs [Eqs.~(\ref{eq analytical formula global
chernoff}), (\ref{eq global stein}), and (\ref{eq global hoeffding})
], we can directly calculate these optimum error exponents. 
As a result, we can derive the lemma. \hfill $\square$

From the above theorem, we describe the asymptotically optimal one-way LOCC protocols as follows:\\
{\it (One-way LOCC protocol)
\begin{enumerate}
 \item For each copy of unknown states, Alice and Bob perform the 
       projections onto the Schmidt basis of $\ket{\Psi}$. 
\item For resulted independent and identical probability distribution, they perform the 
       likelihood-ratio test $\Lambda\left(t\right)$ defined by Eq.(\ref{eq def lrt}). 
\end{enumerate}
}
In the above protocol, $t$ should be chosen depending on the target 
optimal error exponent. To attain $\xi_{\rightarrow}(\rho_{mix}\|\Psi)$, 
$\eta_{\rightarrow }(\epsilon|\rho_{mix}\|\Psi)$, $A_{\rightarrow}(r|\rho_{mix}\|\Psi)$,
and $B_{\rightarrow}(r|\rho_{mix}\|\Psi)$, 
we need to choose $t$ as $1$, $2^{-n\left(\eta_{\rightarrow }(\epsilon|\rho_{mix}\|\Psi)-\delta\right)}$,
$2^{-n\left( A_{\rightarrow}(r|\rho_{mix}\|\Psi)-r \right)}$,
and $2^{n\left( B_{\rightarrow}(r|\rho_{mix}\|\Psi)-r \right)}$, respectively.

The equation $\theta_{\rightarrow}(\epsilon|\rho_{mix}\|\Psi)=+ \infty$ guarantees
that $\beta_{n,\rightarrow}(\epsilon|\rho_{mix}\|\Psi)$ super-exponentially converges to $0$ in the
limit of $n \rightarrow \infty$ for a fixed $\epsilon >0 $.
Indeed, in Section I\hspace{-.1em}I\hspace{-.1em}I of the last paper \cite{OH10},
it was proved that $\beta_{n,C}(\epsilon|\rho_{mix} \|\Psi)=0$ for $C=\rightarrow, \leftrightarrow,
sep$, and $\epsilon \ge 1/\max \left( d_A, d_B \right)$. 
Since $\beta_{n,\rightarrow}(\epsilon|\rho_{mix} \|\Psi) \ge \beta_{n,\leftrightarrow}(\epsilon|\rho_{mix} \|\Psi) \ge
\beta_{n,sep}(\epsilon|\rho_{mix} \|\Psi)$ by the definitions, we derive 
the following corollary:
\begin{Corollary} \label{corollary beta R = infty}
 For $n \ge -\frac{\log \epsilon}{\log \left (\max \left( d_A, d_B \right)\right)}$,
\begin{align}
\beta_{n,\rightarrow}(\epsilon|\rho_{mix}\|\Psi)=\beta_{n,\leftrightarrow}(\epsilon|\rho_{mix}\|\Psi)=
 \beta_{n,sep}(\epsilon|\rho_{mix}\|\Psi)\nonumber \\
=0,
\end{align} 
where $\beta_{n,C}(\epsilon|\rho \|\sigma)$ is defined by Eq.~(\ref{eq def beta n C alpha rho sigma}).
Thus,
\begin{align}
\theta_{\rightarrow}(\epsilon|\rho_{mix}\|\Psi)=\theta_{\leftrightarrow}(\epsilon|\rho_{mix}\|\Psi)=
 \theta_{sep}(\epsilon|\rho_{mix}\|\Psi)\nonumber \\
=+\infty, \label{eq lemma beta R = infty}
\end{align}
where $\theta_{C}(\epsilon|\rho\|\sigma)$ is defined by Eq.~(\ref{eq def stein beta}). 
\end{Corollary}

\section{Hypothesis testing under separable POVMs}\label{sec separable}
In this section, we consider $C=sep$, that is, the local
hypothesis testing under separable LOCC POVMs.

\subsection{Stein's Lemma under separable POVMs}
In this subsection, we treat the optimal error exponents 
$\eta_{sep}\left(\epsilon|\rho_{mix}\|\Psi\right)$ and 
$\theta_{sep}\left(\epsilon|\rho_{mix}\|\Psi\right)$,  which are the optimal
error exponents  in the
 problem setting of Stein's Lemma with an additional separability
 condition on POVMs.

We have derived the equality $\theta_{sep}\left(\epsilon|\rho_{mix}\|\Psi\right)=+\infty$ by means
of the result of one-way LOCC POVMs in Corollary \ref{corollary beta R = infty}.
Hence, we treat the other error exponent, $\eta_{sep}\left(\epsilon|\rho_{mix}\|\Psi\right)$, here.
The analytical formula for this error exponent is given as follows:
\begin{Theorem}\label{theorem sec asymptotic}
\begin{align}
\eta_{\rightarrow}\left(\epsilon|\rho_{mix}\|\Psi\right)=
\eta_{\leftrightarrow}\left(\epsilon|\rho_{mix}\|\Psi\right) = 
\eta_{sep}\left(\epsilon|\rho_{mix}\|\Psi\right) \nonumber 
 \\
= \log d_A + \log d_B - E(\ket{\Psi}), \label{eq sec asymptotic theorem}
\end{align}
where $\eta_{C}\left(\epsilon|\rho_{mix}\|\Psi\right)$ is defined by 
 Eq.~(\ref{eq def stein alpha}), and $E(\ket{\Psi})$ is the entropy of entanglement defined by Eq.~(\ref{eq def entropy of entanglement}). 
\end{Theorem}
From Corollary \ref{corollary beta R = infty} and this theorem,  we
observe the following fact:
Although there is a gap in optimal error probabilities among 
one-way LOCC, two-way LOCC, and separable POVMs in the non-asymptotic
local hypothesis testing \cite{OH10},
such a difference never appears on the Stein's lemma type of error exponents  
in their asymptotic extensions.
In the next subsection, we will see that such a difference appears in the
Chernoff and Hoeffding types of optimal error exponents. 
We also note that this theorem gives a new operational interpretation
of the entropy of entanglement in terms of the local hypothesis testing.

There is a strong mathematical relation between the optimal error 
exponent $\eta_{sep}\left(\epsilon|\rho_{mix}\|\Psi \right)$ and the
environment-assisted capacities of the quantum channel treated in
\cite{W05}. In particular, we can use Lemma 6 of \cite{W05} to derive an
upper bound for $\eta_{sep}\left(\epsilon|\rho_{mix}\|\Psi \right)$. 
However, we give a direct proof of the theorem without using this lemma
in \cite{W05} to derive a slightly stronger result, that is, {\it the
strong converse bound} (Corollary \ref{corollary strong converse}).

({\bf Proof of Theorem \ref{theorem sec asymptotic}})\\
By the definitions, we have
\begin{align}
 \underline{\eta}_{\rightarrow}\left(\epsilon|\rho_{mix}\|\Psi \right) & 
 \le
 \underline{\eta}_{\leftrightarrow}\left(\epsilon|\rho_{mix}\|\Psi 
				  \right) \nonumber \\
&\le 
 \underline{\eta}_{sep}\left(\epsilon|\rho_{mix}\|\Psi \right) \le 
 \overline{\eta}_{sep}\left(\epsilon|\rho_{mix}\|\Psi \right). 
\end{align}
In the above inequalities, $\underline{\eta}_{C}\left(\epsilon|\rho_{mix}\|\Psi \right)$
and $\overline{\eta}_{C}\left(\epsilon|\rho_{mix}\|\Psi \right)$ are 
defined by 
\begin{align}
\underline{\eta}_{C}\left(\epsilon|\rho \|\sigma \right)
\stackrel{\rm def}{=} \underline{\lim}_{n \rightarrow \infty}
 -\frac{1}{n}\log \alpha_{n,C}(\epsilon|\rho\| \sigma),\nonumber \\
\overline{\eta}_{C}\left(\epsilon|\rho \|\sigma \right)
\stackrel{\rm def}{=} \overline{\lim}_{n \rightarrow \infty}
 -\frac{1}{n}\log \alpha_{n,C}(\epsilon|\rho\| \sigma),
\end{align}
where $\underline{\lim}$ and $\overline{\lim}$ are the limit inferior 
and the limit superior, respectively, and 
$\alpha_{n,C}(\epsilon|\rho\| \sigma)$ is defined by Eq.~(\ref{eq def alpha n C alpha rho sigma}).
Thus, from Eq.~(\ref{eq alpha R epsilon rightarrow}), we immediately derive 
\begin{equation}\label{eq sec asymptotic lower bound}
		      \underline{\eta}_{\rightarrow}\left(\epsilon|\rho_{mix}\|\Psi \right) \ge \log d_A +
		     \log d_B - E(\ket{\Psi}).
\end{equation}

Hence, what we need to prove is inequality 
\begin{equation}\label{eq sec asymptotic upper bound}
 \overline{\eta}_{sep}\left(\epsilon|\rho_{mix}\|\Psi \right)
		     \le \log d_A +
		     \log d_B - E(\ket{\Psi}).
\end{equation}
To show the inequality, we define $\widetilde{\eta}_{sep}\left(\rho \|\sigma \right)$ as
\begin{align}
\quad& \widetilde{\eta}_{sep}\left(\rho \|\sigma \right)
 \nonumber \\ 
\stackrel{\rm def}{=}& \sup_{\{T_n\}_{n = 1}^{\infty}} \Big \{ 
\overline{\lim}_{n \rightarrow
 \infty } -\frac{1}{n} \log \alpha_{n}(T_n) \ \Big |
 \nonumber \\
\quad& \qquad \overline{\lim}_{n \rightarrow \infty} \beta_{n}(T_n)
 < 1, \{T_n, I -T_n\}
 \in Sep \Big \}, \label{eq def tilde eta}
\end{align} 
where $\alpha_n(T_n)$ and $\beta_n(T_n)$ are defined by Eqs.~(\ref{eq def 
type 1 error}) and (\ref{eq def type 2 error}), respectively.
By the definition of $\overline{\eta}_{sep}\left(\epsilon|\rho \|\sigma \right)$, there exists a sequence
of separable POVMs $\{T_n \}_{n=1}^\infty$ such that $\overline{\lim}
-\frac{1}{n}\log \alpha_{n}(T_n)=\overline{\eta}_{sep}\left(\epsilon|\rho \|\sigma 
					   \right)$ and $\beta_n(T_n) \le
\epsilon$ for all $n$. This fact and the definition of 
$\widetilde{\eta}_{sep}(\rho \|\sigma)$
guarantee 
\begin{equation}\label{eq widetilde eta sep ge eta sep}
	     \widetilde{\eta}_{sep}\left(\rho \|\sigma \right) \ge \overline{\eta}_{sep}\left(\epsilon|\rho \|\sigma \right)
\end{equation} 
for all $\rho$, $\sigma$, and $0 < \epsilon < 1$. 
Hence, the proof of Eq.~(\ref{eq sec asymptotic upper bound}) is reduced 
to that of the inequality 
\begin{equation}
	 \tilde{\eta}_{sep}\left(\rho_{mix}\|\Psi\right)\le 
	\log d_A + \log d_B -E\left( \ket{\Psi} \right). 
\label{tilde eta le log d a d b e}
\end{equation}

In the following, to prove Eq.~(\ref{eq sec asymptotic upper 
bound}), we show that
when a sequence of tests $\{T_n,I-T_n\}\in Sep$ satisfies 
$\overline{\lim}_{n \rightarrow \infty} \beta_{n}(T_n)<1$,
the following inequality holds:
\begin{align}
\overline{\lim}_{n \rightarrow \infty } -\frac{1}{n} \log 
 \alpha_{n}(T_n) 
\le 
\log d_A + \log d_B - E(\ket{\Psi}),\label{h2}
\end{align}
which is equivalent to the following inequality:
\begin{align}\label{eq underline lim log tr t n ge}
\underline{\lim}_{n \rightarrow \infty } \frac{1}{n} \log 
 \Tr T_n 
\ge E(\ket{\Psi}).
\end{align}

Suppose a sequence of tests $\{T_n,I-T_n\}\in Sep$ satisfies 
\begin{align}
&\overline{\lim}_{n \rightarrow \infty} \beta_{n}(T_n)=1-\underline{\lim}_{n \rightarrow \infty}\bra{\Psi}^{\otimes n}T_n 
\ket{\Psi}^{\otimes n} < 1.
\end{align}
Then, there exist a small real number $\delta>0$ and an integer $N_1$ such that
\begin{align}
   \langle \Psi |^{\otimes n} T_{n}|\Psi\rangle^{\otimes n} > 
 \delta  \label{h1}
\end{align}
for any number $n \ge N_1$.

In the following part, in order to bound $\Tr T_n$ and derive 
Eq.~(\ref{eq underline lim log tr t n ge}), we will construct an  
unnormalized vector $\ket{\Upsilon_n} \in \Hi_{AB}^{\otimes n}$, which is 
close to $\ket{\Psi}^{\otimes n}$ and whose 
largest Schmidt coefficient is on the order of $e^{-n E\left(\ket{\Psi}\right)}$. 

For an arbitrary small real number $\delta'>0$ and the above given $\delta>0$,
there exists an integer $N_2(\delta,\delta')$ such that
$\Tr L_{\delta',n}^A \sigma_A^{\otimes n} \ge 1-\delta^2/16$
where $\sigma_A \stackrel{\rm def}{=} \Tr_B \ket{\Psi}\bra{\Psi}\in\B\left(\Hi_{A}\right)$,
\begin{equation}\label{eq def L delta' n A}
L_{\delta',n}^A:= \{ \sigma_A^{\otimes n} - 
 e^{-n(E(|\Psi\rangle)-\delta')}\le 0 \}\in\B\left(\Hi_{A}^{\otimes n}\right),
\end{equation}
and $\{X \le 0\}$ is the projection to the subspace spanned by eigenvectors with a non-positive eigenvalue of $X$.
$L_{\delta',n}^B\in\B\left(\Hi_{B}^{\otimes n}\right)$ is also defined similarly.
Then, the definitions of $L_{\delta',n}^A$ and $L_{\delta',n}^B$ lead
\begin{align}\label{eq bra psi otimes n l delta n = tr l delta n}
& \bra{\Psi}^{\otimes n} L_{\delta',n}^A\otimes 
 L_{\delta',n}^B \ket{\Psi}^{\otimes n}=\Tr L_{\delta',n}^A \sigma_A^{\otimes n} \ge  1-\delta^2/16.
\end{align}
By means of $L_{\delta',n}^{A(B)}$, we define
the unnormalized vector $|\Upsilon_n\rangle$ as  
\begin{equation}\label{eq def Upsilon n}
 |\Upsilon_n\rangle \stackrel{\rm def}{=}(L_{\delta',n}^A\otimes 
L_{\delta',n}^B)|\Psi\rangle^{\otimes n}.
\end{equation}
Then, we can show that $|\Upsilon_n\rangle$ is close to 
$\ket{\Psi}^{\otimes n}$ for large $n$ as follows:
For $n \ge \max \{N_1,N_2 \}$,
\begin{align}
& \| |\Psi\rangle^{\otimes n} \langle \Psi |^{\otimes n}-\ket{\Upsilon_n}\bra{\Upsilon_n} \|_1\nonumber \\
= & \| |\Psi\rangle^{\otimes n} \langle \Psi |^{\otimes n} \nonumber \\
&- L_{\delta',n}^A\otimes L_{\delta',n}^B 
 |\Psi\rangle^{\otimes n} \langle \Psi |^{\otimes n}  L_{\delta',n}^A\otimes L_{\delta',n}^B \|_1 
 \nonumber \\
\le & \| |\Psi\rangle^{\otimes n} \langle 
 \Psi |^{\otimes n}(I-L_{\delta',n}^A\otimes L_{\delta',n}^B) \|_1  \nonumber \\
&+ \| (I- L_{\delta',n}^A\otimes L_{\delta',n}^B)|\Psi\rangle^{\otimes n} \langle 
 \Psi |^{\otimes n} (L_{\delta',n}^A\otimes L_{\delta',n}^B)\|_1  \nonumber \\
=& \|(I-L_{\delta',n}^A\otimes L_{\delta',n}^B) |\Psi\rangle^{\otimes n}  \|  \nonumber \\
&+ \| (I- L_{\delta',n}^A\otimes L_{\delta',n}^B)|\Psi\rangle^{\otimes n}\|\cdot
\| L_{\delta',n}^A\otimes L_{\delta',n}^B|\Psi\rangle^{\otimes n} \|  \nonumber \\
\le &
2
\sqrt{ \langle \Psi |^{\otimes n} (I- L_{\delta',n}^A\otimes 
 L_{\delta',n}^B)  |\Psi\rangle^{\otimes n}}
\le 
2\sqrt{ \delta^2/16}=\delta/2. \label{eq psi otimes n - upsilon n}
\end{align}
In the above inequalities, we use the equation $\|\ket{\Psi}\bra{\Phi} 
\|_1=\|\ket{\Psi}\|\cdot\|\ket{\Phi}\|$, which holds for all 
unnormalized vectors $\ket{\Psi}$ and $\ket{\Phi}$, in the second equality
and Eq.~(\ref{eq bra psi otimes n l delta n = tr l delta n}) in the third 
 inequality.

Thus, for $n \ge \max \{N_1,N_2 \}$, $|\Upsilon_n\rangle$ satisfies
\begin{align}
& \langle \Upsilon_n | T_{n} |\Upsilon_n\rangle \nonumber \\
\ge& 
\langle \Psi |^{\otimes n} T_{n} |\Psi\rangle^{\otimes n}  - \| |\Psi\rangle^{\otimes n} 
 \langle \Psi |^{\otimes n} -|\Upsilon_n\rangle \langle \Upsilon_n | 
 \|_1 \nonumber \\
> &
\delta -\delta/2 =\delta/2, \label{eq upsilon n t n upsilon n ge psi n t 
 n psi n}
\end{align}
where Eqs.~(\ref{h1}) and (\ref{eq psi otimes n - upsilon n}) are used in the second inequality.
Since the largest Schmidt coefficient of $|\Upsilon _n \rangle $ is no more than 
$e^{-n(E(|\Psi\rangle)-\delta')}$,
any separable state $\tau \in \B\left( \Hi_{AB}^{\otimes n}\right)$ satisfies
\begin{align*}
\langle\Upsilon_n| \tau|\Upsilon_n\rangle \le e^{-n(E(|\Psi\rangle)-\delta')}.
\end{align*}
Because $T_{n}$ has a separable form, the above inequality leads to
\begin{align}\label{upsilon n t n upsilon le e n e - delta}
\bra{\Upsilon_n} T_{n}\ket{\Upsilon_n} \le e^{-n(E(|\Psi\rangle)-\delta')} \Tr  T_{n}.
\end{align}
Thus, for $n \ge \max \{N_1,N_2 \}$, 
Eqs.~(\ref{eq upsilon n t n upsilon n ge psi n t 
 n psi n}) and (\ref{upsilon n t n upsilon le e n e - delta}) lead to
\begin{align}\label{tr t n > delta e n e}
\Tr  T_{n}
>
\delta e^{n(E(|\Psi\rangle)-\delta')} /2,
\end{align}
which implies 
\begin{align}
& \underline{\lim}_{n \rightarrow \infty } \frac{1}{n} \log 
 \Tr T_n > E(\ket{\Psi}) - \delta'.
\end{align}
Since $\delta'>0$ is arbitrary, we obtain Eq.~(\ref{eq underline lim log tr t n ge}).
Thus, Eq.~(\ref{h2}) holds when a sequence of tests $\{T_n,I-T_n\}\in Sep$ satisfies 
$\overline{\lim}_{n \rightarrow \infty} \beta_{n}(T_n)<1$, which
implies Eq.~(\ref{tilde eta le log d a d b e}).
Eqs.~(\ref{eq widetilde eta sep ge eta sep}) and (\ref{tilde eta le log d a d b e})
lead to Eq.~(\ref{eq sec asymptotic upper bound}). 
Finally, Eqs.~(\ref{eq sec asymptotic upper bound}) and (\ref{eq sec 
asymptotic lower bound}) complete the proof of the theorem.
\hfill $\square$

In the last part of this section, we consider the strong converse bound 
$\eta^{\dagger}_{C}(\rho \| \sigma)$ defined by Eq.~(\ref{eq def strong converse eta}). 
As we have seen in Section \ref{sec preliminary}, for global POVMs $C=g$, it is known that
$\eta_{g}\left(\epsilon|\rho\|\sigma 
\right)=\eta^{\dagger}_{g}\left(\rho\|\sigma \right)$ for all 
$\epsilon>0$, $\rho$, and $\sigma$ \cite{ON00}.
Actually, a similar equality holds for $C=\rightarrow,
\leftrightarrow, sep$ when $\rho=\rho_{mix}$ and $\sigma
=\Psi$: 
\begin{Corollary}\label{corollary strong converse}
 For $C=\rightarrow, \leftrightarrow, sep$, 
\begin{equation}
 \eta_{C}(\epsilon|\rho_{mix}\|\Psi)=\eta^{\dagger}_{C}(\rho_{mix}\|\Psi)=\log
  d_A + \log d_B - E(\ket{\Psi}),
\end{equation}
where $\eta_{C}(\epsilon|\rho \|\sigma)$ and $\eta^{\dagger}_{C}(\rho \| 
 \sigma)$ are defined by Eqs.~(\ref{eq def 
 stein alpha}) and (\ref{eq def strong converse eta}), respectively. 
\end{Corollary}
{\bf (Proof)} \\
Since we have proved Theorem
\ref{theorem sec asymptotic}, we only need to prove 
\begin{equation}\label{ineq alpha R dagger sep le}
 \eta^{\dagger}_{sep}\left(\rho_{mix}\|\Psi\right) \le \log d_A + \log d_B -E(\ket{\Psi}).
\end{equation}
Hence, because of the 
definition of $\eta^{\dagger}_{sep}\left(\rho_{mix}\|\Psi\right)$ [see 
Eq.~(\ref{eq def strong converse eta})],   
we need to show that
when a sequence of tests $\{T_n,I-T_n\}\in Sep$ satisfies 
$\underline{\lim}_{n \rightarrow \infty} \beta_{n}(T_n)<1$,
the following inequality holds:
\begin{align}
\underline{\lim}_{n \rightarrow \infty } -\frac{1}{n} \log 
 \alpha_{n}(T_n) 
\le 
\log d_A + \log d_B - E(\ket{\Psi}),\label{h20}
\end{align}

The proof of this inequality is almost the same as the last half of the proof of Theorem \ref{theorem sec asymptotic}.
The difference comes from Eq.~(\ref{h1}). 
The condition $\underline{\lim}_{n \rightarrow \infty} \beta_{n}(T_n)<1$ 
 does not guarantee that Eq.~(\ref{h1}) holds 
for all large $n$ in general. What we can say is that there exist a small 
real number $\delta>0$  and  a subsequence $\{n(k)\}_{k=1}^{\infty}$ such that
\begin{align}
\langle \Psi |^{\otimes n} T_{n(k)}|\Psi\rangle^{\otimes n} > \delta \label{h1 strong converse}
\end{align}
 for any number $k$.
As a result, we derive Eq.~(\ref{tr t n > delta e n e}) 
only for the subsequence. That is, we derive 
\begin{align}\label{eq strong converse alpha n sep = Tr T n / d_A^n d_B^n}
\Tr T_{n(k)} > \delta e^{n(k)(E(|\Psi\rangle)-\delta')}/2
\end{align}
for large $k$. 
This inequality implies for an arbitrary small $\delta'>0$
\begin{align}
&\quad \underline{\lim}_{n \rightarrow
 \infty } -\frac{1}{n} \log \alpha_{n}(T_n) \nonumber \\
&<
\log  d_A + \log  d_B -E(|\Psi\rangle)+\delta',
\end{align}
which implies Eq.~(\ref{ineq alpha R dagger sep le}).
\hfill $\square$

\subsection{Chernoff bound for separable POVM}
In this subsection, we treat the optimal error exponent 
$\xi_{sep}\left(\rho_{mix}\|\Psi \right)$, which is the Chernoff bound with an additional separability
 condition on POVMs.

To derive an analytical formula $\xi_{sep}\left(\rho_{mix}\|\Psi \right)$, we need 
one of the main results of Section V\hspace{-.1em}I of the previous paper
\cite{OH10}, where we proved the equivalence between the
hypothesis testing under separable POVMs and a different type of global hypothesis with
a composite alternative hypothesis.
In the paper \cite{OH10}, we only treated a single-copy case. 
However, all the results in \cite{OH10} 
can be easily extended to the $n$-copy case. 
In the $n$-copy cases,  
the corresponding global hypothesis testing on 
$\left(\mathbb{C}^{d}\right)^{\otimes n}$ is a hypothesis testing 
between an arbitrary pure state $\ket{\psi}^{\otimes n}$ (an alternative 
hypothesis) and a set of states
$\left \{ \ket{\phi_{L}^n} \right \}_{L \in 
\mathbb{Z}^{d^n}_2}$ (a null hypothesis), where $d\stackrel{\rm 
def}{=}\min \left( d_A, d_B\right)$.  Here, $\ket{\phi_{L}^{n}} \in 
 \left(\mathbb{C}^{d}\right)^{\otimes 
n}$ is defined as 
\begin{equation}\label{eq def phi l n}
 \ket{\phi_{L}^n}\stackrel{\rm def}{=}\sum _{J=1}^{d^n}
  (-1)^{L_J} \ket{J}, \quad L \in \mathbb{Z}_2^{d^n},
\end{equation}
where $L_J$ is a $J$th element of $L\in \mathbb{Z}_2^{d^n}$, and $\{ 
\ket{J} \}_{J=1}^{d^n}$ is equivalent to the standard basis $\{ 
\ket{i_1}\otimes \cdots \otimes \ket{i_n} 
\}_{i_1,\dots, i_n}$ of the space $\left(\mathbb{C}^{d}\right)^{\otimes n}$ under the relabeling of the basis vectors.
Then, for a two-valued POVM $\{ S_n, I-S_n \}$ on
$\left(\mathbb{C}^{d}\right)^{\otimes n}$, 
the type-1 error $a_n(S_n)$ and the type-2 error $b_n(S_n)$ are defined
as 
\begin{align}
 a_n(S_n) &\stackrel{\rm def}{=} \max_{L \in \mathbb{Z}_2^{d^n}} \bra{\phi_{L}^n}S_n \ket{\phi_{L}^n }
\label{eq def a n}  \\
b_n(S_n) &\stackrel{\rm def}{=} \bra{\psi}^{\otimes n}I_{d}^n-S_n \ket{\psi
 }^{\otimes n}, \label{eq def b n}
\end{align}
where $I_{d}^n$ is the identity operator on
$(\mathbb{C}^{d})^{\otimes n}$.
The optimal type-2 error under the restriction that the
type-1 error is no more than $a\ge0$ can be written as 
\begin{equation}\label{eq def gamma n a}
 \gamma_n \left( a \Big |\{\ket{\phi_{L}^n}\} \Big\| \ket{\psi} \right ) 
  \stackrel{\rm def}{=} \min_{0 \le S_n \le I_{d }^n} \{ b_n(S_n) |
  a_n(S_n) \le a\}.
\end{equation}
In the following part of this subsection, we often abbreviate $ \gamma_n
\left( a \big|\{\ket{\phi_{\vec{K}}^n}\} \big\| \ket{\psi} \right )$
as $ \gamma_n \left( a \right )$.
Then, in the present notation,
the statement of the theorem can be written as follows: 
\begin{Theorem}[Theorem 5 of \cite{OH10}]\label{theorem previous paper 1}
\begin{equation}
\beta_{n,sep}(\alpha | \rho_{mix}\|\Psi)=
\gamma_n\left( \alpha d_{max}^n \Big |\{\ket{\phi_{L}^n}\}
					  \Big \| \ket{\psi} \right ).
\end{equation}
In the above equation, $\beta_{n,sep}(\alpha | \rho \|\sigma)$ and
$\gamma_n\left( a \Big |\{\ket{\phi_{L}^n}\}
					  \Big \| \ket{\psi} \right )$ 
 are defined by Eqs.~(\ref{eq def beta n C alpha rho sigma}) and 
 (\ref{eq def gamma n a}), respectively. $d_{max}$ is defined as 
\begin{equation}\label{eq def d max}
						    d_{max}\stackrel{\rm def}{=}\max \left( d_A, 
						   d_B \right),
\end{equation}
 and $\ket{\psi}$ is a pure state on 
 $\mathbb{C}^{d_{min}}$ defined 
 as 
\begin{equation}\label{eq def ket psi}
     \ket{\psi} \stackrel{\rm def}{=}\sum_{i=1}
    \sqrt{\lambda_i}\ket{i}
\end{equation} 
by using the Schmidt coefficient $\{ \lambda
 _i \}_{i=1}^{d}$ of $\ket{\Psi}$. 
\end{Theorem}

We define $Q_n(\kappa_0,\kappa_1)$ 
as the optimal mean error probability of the above global hypothesis
testing under a given prior $(\kappa_0,\kappa_1)$: 
\begin{equation}\label{eq def Q_n kappa _0 kappa_1}
 Q_n(\kappa_0,\kappa_1) \stackrel{\rm def}{=} 
\min _{0 \le S_n \le I_{d}^n}\kappa_0
  a_n(S_n)+\kappa_1 b_n(S_n).
\end{equation}
Theorem \ref{theorem previous paper 1} immediately leads the following lemma: 
 \begin{Lemma}\label{lemma separable = global composite 1}
The following two equations hold for all priors $\left(\pi_0, \pi_1 \right)$:
  \begin{align}
 &  P_{n,sep}(\pi_0,\pi_1|\rho_{mix} \|\Psi) \nonumber \\
&= \left(
   \frac{\pi_0}{d_{max}^n}+\pi_1 \right)\cdot Q_n
   (\kappa_0(n),\kappa_1(n)), \label{eq P_n sep = Q_n}\\
& \xi_{sep}(\rho_{mix} \|\Psi) \nonumber 
\\ &= \lim_{n \rightarrow \infty}-\frac{1}{n}\log
   Q_n(\kappa_0(n),\kappa_1(n)) \label{eq xi_sep = chernoff of composite}.
  \end{align}
In the above equations, $P_{n,C}(\pi_0,\pi_1|\rho \|\sigma)$, 
$\xi_{C}(\rho \|\sigma)$,  $d_{max}$, and $Q_n(\kappa_0,\kappa_1)$ are defined by Eqs.~(\ref{eq def p n c pi 0 pi 
  1 rho sigma}),  (\ref{eq def xi C}), (\ref{eq def d max}), and (\ref{eq def Q_n kappa _0 kappa_1}), 
  respectively.
 $\kappa_0(n)$ and $\kappa_1(n)$ are defined as 
\begin{align}
 \kappa_0(n)&\stackrel{\rm def}{=} \left(
   \frac{\pi_0}{d_{max}^n}+\pi_1 \right)^{-1}\cdot
 \frac{\pi_0}{d_{max}^n} \label{eq def kappa 0}\\
 \kappa_1(n)&\stackrel{\rm def}{=} \left(
   \frac{\pi_0}{d_{max}^n}+\pi_1 \right)^{-1}\cdot
 \pi_1. \label{eq def kappa 1}
\end{align}
 \end{Lemma}
In the above lemma, Eq.~(\ref{eq xi_sep = chernoff of composite}) means
that the Chernoff bound under separable POVMs is equal to the exponent of
the optimal mean error probability of the above-mentioned global hypothesis
testing problem with the prior $(\kappa_0(n),\kappa_1(n))$ that
converges to $(0,1)$ in the limit $n \rightarrow \infty$.

{\bf (Proof of Lemma \ref{lemma separable = global composite 1})}\\
Eq.~(\ref{eq xi_sep = chernoff of composite}) is just a direct
consequence of Eq.~(\ref{eq P_n sep = Q_n}).

Eq.~(\ref{eq P_n sep = Q_n}) can be derived as follows:
\begin{align}
& P_{n,sep}(\pi_0,\pi_1|\rho_{mix}\|\Psi)\nonumber \\
=& \inf _{0\le \alpha \le 1} \pi_0 \alpha + \pi_1
 \beta_{n,sep}(\alpha|\rho_{mix}\|\Psi) \nonumber\\
=&\inf _{0\le \alpha \le 1} \pi_0 \alpha + \pi_1
 \gamma_{n}\left (\alpha d_{max}^n\Big |\{\ket{\phi_{L}^n}\}
					  \Big \| \ket{\psi} \right) \nonumber\\
=& \inf _{0\le \alpha' \le d_B^n} \frac{\pi_0}{d_{max}^n} \alpha' + \pi_1
 \gamma_{n}\left (\alpha' \Big |\{\ket{\phi_{L}^n}\}
					  \Big \| \ket{\psi} \right ) \nonumber\\
=& \inf _{0\le \alpha' \le 1} \frac{\pi_0}{d_{max}^n} \alpha' + \pi_1
 \gamma_{n}\left (\alpha'\Big |\{\ket{\phi_{L}^n}\}
					  \Big \| \ket{\psi}  \right) \nonumber\\
=& \left(
   \frac{\pi_0}{d_{max}^n}+\pi_1 \right)\cdot Q_n
   (\kappa_0(n),\kappa_1(n))
\end{align}
In the above of equations, we used Eq.~(\ref{eq P_{n,C} =
inf alpha beta}) in the first and fifth line and  Theorem \ref{theorem previous paper
1} in the second line. We also use the fact that$\gamma_n\left (\alpha'\Big |\{\ket{\phi_{L}^n}\}
					  \Big \| \ket{\psi} \right 
					  )=\gamma_n\left (1 \Big |\{\ket{\phi_{L}^n}\}
					  \Big \| \ket{\psi} \right)$ for
all $\alpha' \ge 1$ in the fourth line. 
\hfill $\square$

To evaluate $\xi_{sep}$, we start from the following lemma:
\begin{Lemma}\label{lemma xi _sep le chernoff of simple}
\begin{align}\label{eq xi _sep le chernoff of simple}
\quad & \xi_{sep}(\rho_{mix} \|\Psi) \nonumber \\
&\le \lim _{n \rightarrow \infty} 
- \frac{1}{n}\log P_{n,g} (\kappa_0(n),\kappa_1(n)|\ \ket{\phi_0} \|
\ \ket{\psi}).
\end{align} 
In the above equation, 
$P_{n,g} (\kappa_0,\kappa_1|\ \rho \|
\ \sigma)$ and $\xi_{sep}(\rho \|\sigma)$ are defined by Eqs.~(\ref{eq def p n c pi 0 pi 
  1 rho sigma}) and  (\ref{eq def xi C}), respectively.
$\ket{\phi_0}$ is a pure state on $\mathbb{C}^{d}$defined as 
\begin{equation}\label{eq def ket phi 0}
 \ket{\phi_0} \stackrel{\rm def}{=} \frac{1}{\sqrt{d}}\sum _{i=1}^{d}\ket{i}.
\end{equation}
\end{Lemma}
{\bf (Proof)}
From the definition of $Q_{n}(\kappa_0(n),\kappa_1(n))$ 
and
$P_{n,g}(\kappa_0(n),\kappa_1(n)|\ \ket{\phi_0} \|  \ket{\psi})$, we derive 
\begin{align}
\quad & Q_{n}(\kappa_0(n),\kappa_1(n)) \nonumber\\ 
& = \min _{0\le S_n \le I} \max_{L\in \mathbb{Z}_2^{d^n}}
 \kappa_0(n) \bra{\phi_L^n}S_n\ket{\phi_L^n}+ \kappa_1(n) b_{n}(S_n)
\nonumber\\
&\ge   \min _{0\le S_n \le I}
 \kappa_0(n) \bra{\phi_0^n}S_n\ket{\phi_0^n}+ \kappa_1(n) b_{n}(S_n) \nonumber\\
&=
P_{n,g}(\kappa_0(n),\kappa_1(n)| \ \ket{\phi_0} \|
 \ \ket{\psi}), \label{eq Q_n kappa_0 kappa_1 ge P_n g kappa_0 kappa_1}
\end{align}
where $0$ in $\ket{\phi_0^n}$ means the zero vector in $\mathbb{Z}_2^{d^n}$.
To derive the first line of the above inequalities, we use Eq.~(\ref{eq def Q_n kappa _0 kappa_1}). The last equality can be
derived by using the fact $\ket{\phi_{0}^n}=\ket{\phi_0}^{\otimes
n}$. 
Eqs.~(\ref{eq Q_n kappa_0 kappa_1 ge P_n g kappa_0 kappa_1}) and (\ref{eq xi_sep = chernoff of composite})
guarantee Eq.~(\ref{eq xi _sep le chernoff of simple}) \hfill $\square$

The following lemma gives an analytical formula for the the
right-hand-side of  Eq.~(\ref{eq xi _sep le chernoff of simple}): 
\begin{Lemma}\label{lemma -frac 1 n log P_ n g}
 \begin{align}
&\quad \lim_{n \rightarrow \infty} - \frac{1}{n}\log P_{n,g} (\kappa_0(n),\kappa_1(n)|\ \ket{\phi_0} \|
\ \ket{\psi}) \nonumber \\
&= -2 \log \braket{\psi}{\phi_0} + \log d_{max} \label{eq -frac 1 n log P_ n
  g },
 \end{align}
where
$P_{n,g} (\kappa_0,\kappa_1|\ \rho \|
\ \sigma)$, $d_{max}$, $\kappa_0(n)$, and $\kappa_1(n)$ are defined by Eqs.~(\ref{eq def p n c pi 0 pi 
  1 rho sigma}), (\ref{eq def d max}), (\ref{eq def kappa 0}), and 
 (\ref{eq def kappa 1}), respectively.
\end{Lemma}
{\bf (Proof)}\\
Since we only need to consider the case when $n$ is large, 
we assume $\braket{\psi}{\phi_0}^{2n}\le \frac{1}{2}$.
$P_{n,g} (\kappa_0(n),\kappa_1(n)|\ \ket{\phi_0} \|\ \ket{\psi})$
can be written as 
\begin{align}
&\quad  P_{n,g} (\kappa_0(n),\kappa_1(n)|\  \ket{\phi_0} \|
\ \ket{\psi}) \nonumber \\
&=\frac{1}{2} - \frac{1}{2} \left\| \kappa_0(n)
 \ket{\phi_0}^{\otimes n}\bra{\phi_0}^{\otimes n} - \kappa_1(n)
 \ket{\psi}^{\otimes n}\bra{\psi}^{\otimes n}\right\|_1 \nonumber \\ 
&= \frac{1}{2} - \frac{1}{2}\sqrt{1-\nu_n}, \label{eq P_n g = -2log braket}
\end{align} 
where $\nu_n$ is defined as $ \nu_n \stackrel{\rm def}{=} 4 \braket{\psi}{\phi_0}^{2n}
  \kappa_0(n) \kappa_1(n)$.
To derive the first equality of the above equations, we use the well-known
  Holevo-Helstrom's formula for the optimal discrimination of two
  quantum states   \cite{Helstrom76,Holevo82}.
The second equality is the consequence of the direct calculation of the
  trace norm under the condition $\braket{\psi}{\phi_0}^{2n}\le \frac{1}{2}$.
Finally, we derive Eq.~(\ref{eq -frac 1 n log P_ n g }) as follows:
\begin{align*}
&\quad  \lim _{n \rightarrow \infty}- \frac{1}{n}\log P_{n,g} (\kappa_0(n),\kappa_1(n)| \ket{\phi_0} \|
\ \ket{\psi}) \\
&=\lim_{n \rightarrow \infty} -\frac{1}{n} \left[ \log \nu_n +
 \log(1+o(\nu_n)) -\log 4 \right]\\
&=\lim_{n \rightarrow \infty} -\frac{1}{n}  \log \nu_n = -2 \log \braket{\psi}{\phi_0} + \log d_{max},
\end{align*}
where we use Eq.~(\ref{eq P_n g = -2log braket}) in the second line, and 
the fact that $\lim_{n\rightarrow \infty} \nu _n=0$ in the third
line. \hfill $\square$

Finally, we derive an analytical formula for 
$\xi_{sep}\left(\rho_{mix}\|\Psi \right)$ as follows: 
\begin{Theorem}\label{theorem xi _sep le chernoff of simple}
\begin{align}
\xi_{sep}(\rho_{mix} \|\Psi) &=  -2 \log \braket{\psi}{\phi_0} + \log d_{max} \\
&= \log d_A + \log d_B -LR(\ket{\Psi}). 
\label{eq xi sep = log d A log dB -LR}
\end{align}
In the above equation, $\xi_{sep}(\rho \|\sigma)$, $d_{max}$, and $\ket{\phi_0}$ are 
 defined by Eqs.~(\ref{eq def xi C}), (\ref{eq def d max}), and (\ref{eq 
 def ket phi 0}), respectively.
 $LR(\ket{\Psi})$ is the logarithmic robustness of entanglement \cite{Brandao05,Datta09}, 
defined for a pure state $\ket{\Psi}$ whose Schmidt coefficients are 
 $\left\{\lambda_i \right\}_{i=1}^d$ as \cite{VT99}
\begin{equation}\label{eq def LR}
LR(\ket{\Psi})= 2 \log \sum_{i=1}^{d} \sqrt{\lambda_i},
\end{equation} 
where $d$ is defined as Eq.~(\ref{eq def d}).
\end{Theorem}
We note that this theorem gives a new operational interpretation of the logarithmic
robustness of entanglement in terms of the local hypothesis testing. 

{\bf (Proof)}\\
Lemma \ref{lemma -frac 1 n log P_ n g} and Lemma \ref{lemma xi _sep le
chernoff of simple} guarantee
\begin{equation}
  \xi_{sep} \left(\rho_{mix}\|\Psi\right)\le  -2 \log \braket{\psi}{\phi_0} + \log d_{max}.
\end{equation} 
Thus, we only need to show 
\begin{equation}\label{eq xi sep ge - 2 log braket psi phi_0}
  \xi_{sep} \left(\rho_{mix}\|\Psi\right) \ge  -2 \log \braket{\psi}{\phi_0} + \log d_{max}.
\end{equation} 
The definition of $b_n\left(S_n\right)$ [see Eq.~(\ref{eq def b n}))] 
leads to
\begin{equation}\label{eq b_n S_n,0=0}
b_n(\psi^{\otimes n})=0,
\end{equation} 
where $\psi\stackrel{\rm def}{=}\ket{\psi}\bra{\psi}$. 
On the other hand, 
\begin{align} \label{eq a_n S_n,0}
 a_{n}(\psi^{\otimes n}) =& \max_{L \in \mathbb{Z}_2^{d^n}} \bra{\phi_L^n}\psi^{\otimes n} 
 \ket{\phi_L^n} \nonumber \\
=& \bra{\phi_0^n}\psi^{\otimes n} 
 \ket{\phi_0^n} = \braket{\psi}{\phi_0}^{2n},
\end{align}
In the above equations,  
the inequality
$\bra{\phi_L^n}\psi^{\otimes n}\ket{\phi_L^n}\le 
\bra{\phi_0^n}\psi^{\otimes n}\ket{\phi_0^n}$ is used to derive the second equality.
Finally, we can derive inequality (\ref{eq xi sep ge - 2 log braket
psi phi_0}) as follows: 
\begin{align}
&\quad  \xi_{sep}\left(\rho_{mix}\|\Psi\right) \nonumber \\
 &=\lim _{n \rightarrow \infty} \max_{0\le S_n \le I} -\frac{1}{n}\log 
 \left(\kappa_0 (n) a_n(S_n) + \kappa_1(n) b_n(S_n)\right)
 \nonumber \\
& \ge \lim _{n \rightarrow \infty} -\frac{1}{n}\log 
 \left(\kappa_0 (n) a_n(\psi^{\otimes n}) + \kappa_1(n) b_n(\psi^{\otimes n})\right)
 \nonumber \\
&= -2 \log \braket{\psi}{\phi_0} + \log d_{max},
\end{align}
where we use Lemma \ref{lemma separable = global composite 1} to derive the 
first equality and Eqs.~(\ref{eq b_n S_n,0=0}) and (\ref{eq a_n S_n,0}) 
to derive the second equality. 
\hfill $\square$

\subsection{Hoeffding bound for separable POVMs}
In this subsection, we analyze the Hoeffding bound of our hypothesis
testing under separable POVMs. 

Similar to the case of the Chernoff bound, we utilize Theorem
\ref{theorem previous paper 1} to derive a relationship between the
hypothesis testing under separable POVMs and the global hypothesis
testing with a composite alternative hypothesis. 
We define the Hoeffding bound of the corresponding global hypothesis testing
as 
\begin{align}
 &\quad \Gamma\left (r\Big |\{\ket{\phi_{L}^n}\} \Big\| \ket{\psi}\right 
 ) 
\nonumber \\
& \stackrel{\rm def}{=} \sup _{\{S_n\}_{n=1}^{\infty}} \{ \underline{\lim}_{n
  \rightarrow \infty} -\frac{1}{n}\log b_n(S_n) |  \nonumber \\
& \quad  \quad \underline{\lim}_{n
  \rightarrow \infty} -\frac{1}{n}\log a_n(S_n) \ge r, \  0 \le S_n \le
 I_{d}^n  \}, \label{eq def gamma r big}
\end{align}
where $\ket{\phi_L^n}$, $a_n(S_n)$ and $b_n(S_n)$ are defined by Eqs.~(\ref{eq def phi l 
n}), (\ref{eq def a n}) and (\ref{eq def b n}).
Then, Theorem \ref{theorem previous paper 1} is immediately rewritten as 
follows:
\begin{Lemma}\label{lemma B_sep r = b^R r - log d}
 For $r \ge \log d_{max}$, 
\begin{align}
 B_{sep}(r|\rho_{mix}\|\Psi)&= \Gamma \left ( r-\log d_{max}\Big |\{\ket{\phi_{L}^n}\} \Big\| \ket{\psi}\right),
\end{align}
where $B_{sep}(r|\rho\|\sigma)$, $d_{max}$, and $\Gamma \left ( r\Big 
 |\{\ket{\phi_{L}^n}\} \Big\| \ket{\psi}\right)$ are defined by 
 Eqs.~(\ref{eq def b c}), (\ref{eq def d max}) and (\ref{eq def gamma r 
 big}), respectively.
\end{Lemma}
On the other hand, we can evaluate $\Gamma\left (r\Big |\{\ket{\phi_{L}^n}\} \Big\| \ket{\psi}\right 
 ) $ as follows: 
\begin{Lemma}\label{lemma b^R r}
\begin{align}
& \Gamma\left (r\Big |\{\ket{\phi_{L}^n}\} \Big\| \ket{\psi}\right 
 ) = + \infty, \nonumber \\
& \quad \mbox{\rm if} \ 0 \le r \le -\log
 |\braket{\psi}{\phi_0}|^2  \label{eq b^R r = infty}.\\
 & \Gamma\left (r\Big |\{\ket{\phi_{L}^n}\} \Big\| \ket{\psi}\right 
 )  \le - \log |\braket{\psi}{\phi_0}|^2, \nonumber \\
&\quad \mbox{\rm if} \ r >
 -\log
 |\braket{\psi}{\phi_0}|^2 \label{eq b^R r le -log braket},
\end{align} 
where $\Gamma \left ( r\Big 
 |\{\ket{\phi_{L}^n}\} \Big\| \ket{\psi}\right)$ are defined by 
 Eq.~(\ref{eq def gamma r 
 big}). 
\end{Lemma}
{\bf (Proof)}\\
First, we
observe $\underline{\lim} -
\frac{1}{n}\log a_n(\psi^{\otimes n})=-\log|\braket{\psi}{\phi_0}|^2$ and
$\underline{\lim}-\frac{1}{n} \log b_n(\psi^{\otimes n})=+\infty$, where 
$\psi \stackrel{\rm def}{=}\ket{\psi}\bra{\psi}$. These equations
guarantee Eq.~(\ref{eq b^R r = infty}).

Second, we will prove Eq.~(\ref{eq b^R r le -log braket}). Suppose
$\Gamma_0(r)$ is defined as 
\begin{align}
&\quad \Gamma_0(r) \nonumber \\
& \stackrel{\rm def}{=} \sup _{\{S_n\}_{n=1}^{\infty}} \{ \underline{\lim}_{n
  \rightarrow \infty} -\frac{1}{n}\log b_n(S_n) |  \nonumber \\
& \quad  \quad \underline{\lim}_{n
  \rightarrow \infty} -\frac{1}{n}\log \bra{\phi_0}^{\otimes n}S_n
\ket{\phi_0}^{\otimes n} \ge r, \  0 \le S_n \le I_{d^n}  \}. 
\label{eq def gamma  0 r}
\end{align}
Then, the inequality \begin{align}
\bra{\phi_0}^{\otimes n}S_n
\ket{\phi_0}^{\otimes n} \le \max_{L \in \mathbb{Z}_2^{d^n}} \bra{\phi_{L}^n}S_n \ket{\phi_{L}^n }=a_{n}(S_n)
\end{align} 
guarantees
\begin{equation}\label{eq b^R le b^R_0}
 \Gamma\left (r\Big |\{\ket{\phi_{L}^n}\} \Big\| \ket{\psi}\right 
 )  \le \Gamma_0(r).
\end{equation}
On the other hand, by the definition [see Eq.~(\ref{eq def gamma  0 r})], we observe $\Gamma_0(r)=B_g(r|
\ \ket{\phi_0}  \| \ \ket{\psi})$. This and Eq.~(\ref{eq global hoeffding}) guarantee
\begin{align}
& \Gamma(r) \nonumber \\
&=\left \{
\begin{array}{l}
+\infty,  \quad {\rm if} \ 0 \le r \le  -\log |\braket{\psi}{\phi_0}|^2,
\\
\quad  \\
 -\log
 |\braket{\psi}{\phi_0}|^2, \quad  {\rm if} \ r \ge  -\log |\braket{\psi}{\phi_0}|^2. \\
\end{array}
\right .
\end{align}
This equation together with Eq.~(\ref{eq b^R le b^R_0}) guarantees Eq.~(\ref{eq b^R r le -log braket}). 
\hfill $\square$

The Stein-type's strong converse bound $\eta^{\dagger} _{sep}\left(\rho_{mix}\|\Psi\right)$ also gives
information about the value of $B_{sep}(r|\rho_{mix}\|\Psi)$ as follows: 
\begin{Lemma}\label{lemma strong converse to hoeffding sep}
 For $r > \eta^{\dagger}_{sep}\left(\rho_{mix}\|\Psi \right)$, $B_{sep}(r|\rho_{mix}\|\Psi)=0$,
\end{Lemma}
where $\eta^{\dagger}_{sep}\left(\rho\|\sigma \right)$ and 
$B_{sep}(r|\rho\|\sigma)$ are defined by Eqs.~(\ref{eq def strong converse})
 and (\ref{eq def b c}), respectively.\\
{\bf (Proof)} \\
Suppose a sequence of separable POVMs $\{T_n\}_{n=1}^{\infty}$ satisfies
$\underline{\lim} -\frac{1}{n}\log \alpha(T_n) > \eta^{\dagger}_{sep}\left(\rho\|\sigma \right)$.
Then, from the definition of $\eta^{\dagger}_{sep}\left(\rho\|\sigma 
\right)$ [see Eq.~(\ref{eq def strong converse})], $\lim
-\frac{1}{n}\log \beta_n(T_n)=0$. 
This fact guarantees the statement of the lemma.
\hfill $\square$

Finally, we evaluate $B_{sep}(r|\rho_{mix}\|\Psi)$ as follows: 
\begin{Theorem}\label{theorem capital b_sep r}
\begin{align}
&\quad B_{sep}(r|\rho_{mix}\|\Psi)  \nonumber \\
&=\left \{
\begin{array}{l}
+\infty,  \quad {\rm if} \ 0 \le r \le \log d_A + \log d_B - LR(\ket{\Psi}),
\\
\quad  \\
 0, \quad  {\rm if} \ r \ge  \log d_A + \log
 d_B -E(\ket{\Psi}).\\
\end{array}
\right . \label{B_sep r = + infty}\\
&\quad  B_{sep}(r|\rho_{mix}\|\Psi) \le \log d- LR(\ket{\Psi}), \nonumber \\
& \qquad \mbox{\rm if} \ r >
\log d_A + \log d_B - LR(\ket{\Psi})  \label{eq B sep r le -log braket}.
\end{align}  
In the above equation, $d$ and  $B_{sep}(r|\rho\|\sigma)$  are defined by 
 Eqs.~(\ref{eq def d}) and (\ref{eq def b c}), respectively.
$E(\ket{\Psi})$ and $LR(\ket{\Psi})$ are the entropy of entanglement and 
 the logarithmic robustness of entanglement defined by Eqs.~(\ref{eq def entropy of 
 entanglement}) and (\ref{eq def LR}), respectively.
\end{Theorem}
{\bf (Proof)}\\
Lemmas \ref{lemma B_sep r = b^R
r - log d}, \ref{lemma b^R r}, and
 \ref{lemma strong converse to hoeffding sep} and Corollary 
 \ref{corollary strong converse} guarantee this theorem.
\hfill $\square$

From this theorem, we can also evaluate $A_{sep}(r|\rho_{mix}\|\Psi)$: 
\begin{Corollary}\label{corollary A sep r} 
For $\ r \ge \log d - LR(\ket{\Psi})$, $A_{sep}(r|\rho_{mix}\|\Psi)$ is given as
\begin{align}\label{eq A sep r = log d max - log braket}
 A_{sep}(r|\rho_{mix}\|\Psi)= \log d_A + \log d_B - LR(\ket{\Psi}),
\end{align}
and for $0 \le r \le \log d - LR(\ket{\Psi})$, it is evaluated as 
\begin{align}
&\quad \log d_A +\log d_B - LR(\ket{\Psi}) \nonumber \\
&\le A_{sep}(r|\rho_{mix}\|\Psi) \le \log
  d_A + \log d_B -E(\ket{\psi}). \label{eq log d max - log braket le A
  sep r le}
\end{align}
In the above equation, $d$ and  $B_{sep}(r|\rho\|\sigma)$  are defined by 
 Eqs.~(\ref{eq def d}) and (\ref{eq def a c}), respectively.
$E(\ket{\Psi})$ and $LR(\ket{\Psi})$ are the entropy of entanglement and 
 the logarithmic robustness of entanglement defined by Eqs.~(\ref{eq def entropy of 
 entanglement}) and (\ref{eq def LR}), respectively.
\end{Corollary}
{\bf (Proof)}\\
First, the second inequality of Eq.~(\ref{eq log d max - log braket le A
  sep r le}) comes from the inequality $A_{sep}(r|\rho_{mix}\|\Psi) \le \eta{\dagger}_{sep}(\rho_{mix}\|\Psi)$.
Second, Eq.~(\ref{B_sep r = + infty}) guarantees 
\begin{equation}\label{eq A sep r ge log d max -log braket}
					  A_{sep}(r|\rho_{mix}\|\Psi)\ge 
\log d_A + \log d_B -LR(\ket{\Psi}),			
\ \forall r>0.	
\end{equation} 
This inequality guarantees the first inequality of Eq.~(\ref{eq log d max - log braket le A
  sep r le}). 
Finally,  Eq.~(\ref{eq B sep r le -log braket}) guarantees that there is
  no sequence of separable POVMs $\{ T_n \}_{n=1}^{\infty}$ satisfying
  $\underline{\lim}-\frac{1}{n}\log \alpha_n(T_n) > \log d_A + \log d_B -LR(\ket{\Psi})$ and  
 $\underline{\lim}-\frac{1}{n}\log \beta_n(T_n) > \log d-LR(\ket{\Psi})$.
This fact and the inequality (\ref{eq A sep r ge log d max -log braket}) guarantee
the equality (\ref{eq A sep r = log d max - log braket}).
\hfill $\square$

Here, note that when $\ket{\Psi}$ is a maximally entangled state, we
have $\log d_A + \log d_B - LR(\ket{\Psi})=\log d_A +\log d_B
-E(\ket{\Psi})= \log d_{max}$ and $\log d- LR(\ket{\Psi})=0$. 
Hence, in this case, Theorem \ref{theorem capital b_sep r} and Corollary 
\ref{corollary A sep r} completely
determine the behavior of $B_{sep}(r|\rho_{mix}\|\Psi)$ and $A_{sep}(r|\rho_{mix}\|\Psi)$, respectively.

\section{Hypothesis testing under two-way LOCC POVM} \label{sec two-way}
\subsection{three-step LOCC protocol}

In this section, we consider $C=\leftrightarrow$, that is, the local
hypothesis testing under two-way LOCC POVMs. 

In the previous paper \cite{OH10}, we showed that there is a gap between
the optimal error probabilities under one-way and two-way LOCC POVMs in
non-asymptotic problem settings of this local hypothesis testing.  
On the other hand, in asymptotic problem settings, we have already proved that
there is no difference among one-way, two-way LOCC, and
separable POVMs
in terms of the Stein's lemma type of the optimal error exponents
$\eta_{C}\left(\epsilon|\rho_{mix}\|\Psi \right)$ and $\theta_{C}\left(\epsilon|\rho_{mix}\|\Psi \right)$ in Theorem \ref{theorem sec asymptotic}.
Hence, in this section, we concentrate on the Chernoff and Hoeffding bounds
under two-way LOCC POVMs, and ask the question of whether there is a gap
between one-way and two-way LOCC POVMs
even in the asymptotic setting. 

Since the definition of the two-way
LOCC is mathematically complicated in comparison to that
of one-way LOCC and separable operations, it seems impossible to derive
an analytical formula for the Chernoff and Hoeffding bounds under two-way
LOCC POVM. Hence, we try to derive a lower bound of the Chernoff and
Hoeffding bounds under two-way LOCC POVM by numerical calculations.

Since an infinite number of parameters is necessary to describe general
two-way LOCC operations,
it is impossible to optimize the error exponent numerically for general
two-way LOCC protocols. Thus, we consider only a special
class of three-step LOCC protocols, that can outperform any one-way LOCC
protocols in non-asymptotic problem settings \cite{OH10,OH08}.
However, even if we restrict ourselves to
the special class of three-step LOCC protocols, 
we need to treat infinitely many parameters in the asymptotic situation. 
To simplify the analysis further, 
we consider only the situation 
where Alice and Bob apply the same three-step
LOCC protocols to each single copy of the unknown state and then apply the optimal classical testing 
between two classical hypotheses, i.e., 
two probability distributions given by the above three-step LOCC protocol
and two candidate states.
In other words, we calculate the Chernoff and Hoeffding bounds 
for this particular class of two-way LOCC protocols, 
which are, by the definitions, the lower bounds of the Chernoff and Hoeffding bounds 
for (general) two-way LOCC POVMs.    \\
{\it (Intuitive description of our three-step protocol) \\

\quad The most salient characteristic of 
our three-step LOCC protocol is
Bob's measurement which 
is mutually unbiased for the eigenbasis of Bob's 
local state when the unknown state is  $\ket{\Psi}$.
As a result, Bob's measurement result is completely random,
and Bob's role is to send his state to Alice in a quantum manner. 
It is known that one party can teleport his local state of 
joint state $\ket{\Psi}$ to another party
by the measurement in the mutually unbiased basis;
this procedure gives one of the optimal local discrimination protocols 
for a simultaneously Schmidt decomposable set of states \cite{VSPM01,OH06}. 
Now, our candidate states are not simultaneously Schmidt decomposable,
and thus, Bob cannot teleport his local state perfectly to Alice.
However, our previous results \cite{OH08,OH11} and the following calculation show that the mutually unbiased 
measurement is useful for our problem setting, too.  
Alice's first and second measurements are chosen so as to 
be convenient for numerical optimization.}

To describe the detail of our three-step LOCC protocol for one copy,
we prepare notations.
Let $\mathcal{P}(d_A)$ be
the power set (a set of all subsets) of a finite set $\{ 1, \dots, d_A\}$ excluding an empty set $\emptyset$.
Then, we fix a collection $
{\bf m}:=\{m_\omega\}_{\omega \in \mathcal{P}(d_A)}$ 
of non-negative measures on $\{ 1, \dots, d_A\}$
satisfying the following conditions;
(1) $\sum_{\omega \in \mathcal{P}(d_A)}m_\omega(i) \le 1$ 
for any $i\in \omega$.
(2) The support of $m_\omega$ is $\omega \subset \{ 1, \dots, d_A\}$.
Our three-step LOCC protocol for one copy is given as follows:\\
{\it (Three-step LOCC protocol for one copy)
\begin{enumerate}
\item Alice measures her state by a POVM $\{ M_\omega \}_{\omega \in
       \mathcal{P}(d_A)}$ defined by
\begin{equation}\label{eq def M omega}
		M_\omega \stackrel{\rm def}{=}
	\sum_{i\in \omega} m_{\omega}(i) \ket{i}\bra{i} 
\end{equation} 
where $\{ \ket{i}
       \}_{i=1}^{d_A}$ is Alice's part of the Schmidt basis of $\ket{\Psi}$.

%       Then,  $M_\emptyset$ is defined as 
%       $M_\emptyset \stackrel{\rm def}{=}I_A-\sum_{\omega \in        \mathcal{P}(d_A)/\emptyset}M_\omega$,
%where $\mathcal{P}(d_A)/\emptyset$ denotes a set of all non-empty subset of $\{ 1, \dots, d_A\}$.
%       When Alice's measurement outcome is $\omega=\emptyset$, Alice and Bob stop the protocol and
%       conclude the unknown state to be $\rho_{mix}$. Otherwise, they
%       continue the protocol.

\item Bob measures his state by a POVM 
       $\{ N_j^\omega\}_{j=0}^{|\omega|}$ depending on Alice's
      measurement outcome $\omega$, where $|\omega|$ is a size of set 
       $\omega\subset \{1, \dots, d_A \}$. 
      For $j \in \{1, \dots, |\omega| \}$, $N_j^\omega$ is defined as
      $N_j^\omega=\ket{\xi_j^\omega}\bra{\xi_j^\omega}$, where
      $\{\ket{\xi_j^\omega}\}_{j=1}^{|\omega|}$ is a mutually unbiased basis of the
      subspace $span\{\ket{h} \}_{h \in \omega}$. Then, $N_0^\omega$ is defined as
      $N_0^\omega \stackrel{\rm def}{=} I_B - \sum_{j=1}^{|\omega|}
      N_j^\omega$.
      When Bob derives the measurement outcome $j=0$, Alice and Bob stop 
       the protocol. Otherwise, they 
      continue the protocol.

\item Alice measures her states by a two-valued POVM $\{
      O^{\omega j}_k \}_{k\in \{0,1\}}$, which is defined as follows.
The POVM element $O^{\omega j}_0$ is chosen as Alice's state after the Bob's measurement
      when the given state is $\ket{\Psi}$.
      Hence, $O^{\omega j}_0$ is defined as
      \begin{equation}\label{eq def O omega j}
       O^{\omega j}_0 \stackrel{\rm def}{=} \frac{\sqrt{M_\omega
      \sigma_A}\left(\ket{\xi_j^\omega}\bra{\xi_j^\omega} \right)^T \sqrt{M_\omega
      \sigma_A}}{\bra{\xi_j^\omega} M_\omega \sigma_A
      \ket{\xi_j^\omega}},
      \end{equation}
      where $\sigma_A \stackrel{\rm def}{=}\Tr _B \ket{\Psi}\bra{\Psi}$ 
      and $T$ is the transposition in the Schmidt basis of $\ket{\Psi}$.  
Then, the other POVM element
$O^{\omega j}_1$ is defined as $O^{\omega j}_1 
       \stackrel{\rm def}{=}I_A-O^{\omega j}_0$. 
\end{enumerate}
}
Here, note that 
the free parameters in the three-step LOCC protocol
are only the diagonal elements of
the first Alice's POVM 
$\{ M_\omega \}_{\omega \in \mathcal{P}(d_A)}$, which is characterized by
the collection ${\bf m}:=\{m_\omega\}_{\omega \in \mathcal{P}(d_A)}$ 
of non-negative measures.
In the following, we denote the set of outcomes 
$\delta=\left( \omega, j, k \right)$ by $\Delta$.
Then, we write the output distribution as $P(\delta|\rho_{in}, {\bf m})$
when the initial state is $\rho_{in}=\Psi$ or $\rho_{mix}$
and the applied protocol is characterized by ${\bf m}$.

From the above description of the protocol, 
we can calculate $P(\delta|\rho_{in}, {\bf m})$ as follows: 
The support of $P(\delta|\Psi, {\bf m})$ consists of 
outcomes
$\delta=\left(\omega,j,k \right)$ satisfying 
$\omega \in \mathcal{P}(d_A)$, $1 \le j \le |\omega|$, and $k=0$. 
With this support, $P(\delta|\Psi, {\bf m})$ does not depend on $j$ 
and is given as 
\begin{equation}\label{eq def P delta Psi m omega}
 P\left (\delta=(\omega,j,0)|\Psi, {\bf m} \right )=\frac{1}{|\omega|}\sum _{i\in \omega}m_\omega\left(i \right) \lambda_i.
\end{equation} 
The support of $P(\delta|\rho_{mix}, {\bf m})$ consists of all possible 
$\delta$. 
However, we only give values of $P(\delta|\rho_{mix}, {\bf m})$
on the support of $P(\Psi|\rho_{in}, {\bf m})$
because we use only the values of $P(\delta|\rho_{mix}, {\bf m})$ 
on the support of $P(\delta|\Psi, {\bf m})$ in the following calculation. 
When $\delta=\left(\omega,j,k \right)$ satisfies 
$\omega \in \mathcal{P}(d_A)$, $1 \le j \le |\omega|$, and $k=0$, 
$P(\delta|\rho_{mix}, {\bf m})$ does not depend on $j$
and is given as 
\begin{equation}\label{eq def P delta rho_mix m omega}
 P\left (\delta=(\omega,j,0)|\rho_{mix}, {\bf m} \right )=\frac{\sum 
  _{i\in \omega}\left( m_\omega\left(i \right) \right)^2 \lambda_i}{
d_Ad_B \sum 
  _{i\in \omega} m_\omega\left(i \right) \lambda_i}.
\end{equation} 

Based on the above three-step LOCC protocol for one-copy, 
our two-way LOCC protocol for $n$-copy is described as follows:\\
{\it (Two-way LOCC protocol for $n$-copies)
\begin{enumerate}
 \item Alice and Bob apply the three-step LOCC protocol characterized by ${\bf m}$
to each copy of the unknown state, and obtain the data $(\delta_1, 
       \ldots, \delta_n) \in \Delta^n$.
\item They apply the likelihood-ratio test $\Lambda \left(t\right)$
defined by Eq.(\ref{eq def lrt}) to the resulted $n$-fold independent and identical distribution of 
$P\left (\delta|\rho_{in}, {\bf m} \right )$ on 
$(\delta_1, 
       \ldots, \delta_n) \in \Delta^n$
\end{enumerate}   
}In the above protocol, the parameter $t$ is chosen 
so that the error exponent attains
the Chernoff and Hoeffding bounds
for the classical hypothesis testing between 
$P(\delta|\Psi,{\bf m})$ and $P(\delta|\rho_{mix},{\bf m})$;
thus, $t=1$ for the Chernoff bound, and 
$t=2^{-n\left(A_c(r|P(\delta|\Psi,{\bf m}) \|P(\delta|\rho_{mix},{\bf m}))-r \right)}$
for the Hoeffding bound,
where $A_c(r|p \|q)$ is the classical Hoeffding bound defined 
by Eq.(\ref{eq def xi c A c}).
Finally, the Chernoff and Hoeffding types of error exponents
for the above two-way LOCC protocol for $n$-copies
asymptotically achieve 
the classical Chernoff and Hoeffding bounds between 
$P(\delta|\Psi,{\bf m})$ and $P(\delta|\rho_{mix},{\bf m})$, respectively. 

Optimizing the classical Chernoff and Hoeffding bounds over the 
remaining parameter ${\bf m}$, we derive the
best error exponents that can be achieved by the above two-way LOCC
protocols.
 We write such bounds as 
 $\tilde{\xi}_{\leftrightarrow}\left(\rho_{mix}\| \Psi \right)$ and 
 $\tilde{A}_{\leftrightarrow}\left(r | \rho_{mix} \| \Psi \right)$, where 
$\tilde{\xi}_{\leftrightarrow}\left(\rho_{mix} \| \Psi \right)$ is the Chernoff type and
 $\tilde{A}_{\leftrightarrow}\left(r | \rho_{mix} \| \Psi \right)$ is 
 the Hoeffding type. Hence, they are mathematically defined as  
\begin{align}
& \tilde{\xi} _{\leftrightarrow}\left(\rho_{mix} \| \Psi \right) \nonumber \\
\stackrel{\rm def}{=}& \sup_{  {\bf m}} 
\Big \{   \xi_c \left(P(\delta|\rho_{mix},{\bf m})\|
				    P(\delta|\Psi,{\bf m})  \right) \Big |
 \nonumber \\
& \qquad 0 \le m_\omega(i) \le 1, \sum_{\omega \in O_i} m_\omega(i) \le 1
 \Big \} \label{eq def tilde xi}
\\
&\tilde{A}_{\leftrightarrow}\left(r | \rho_{mix} \| \Psi \right) 
 \nonumber \\ \stackrel{\rm def}{=}& \sup_{ {\bf m}} 
\Big \{A_c \left(r|P(\delta|\rho_{mix},{\bf m})\|
				    P(\delta|\Psi,{\bf m}) \right)
\Big |
 \nonumber \\
& \qquad 0 \le m_\omega(i) \le 1, \sum_{\omega \in O_i} m_\omega(i) \le 1
 \Big \}.\label{eq def tilde A}
\end{align}
In the above formula, $O_i $ is a 
subset of the power set of $\{1,\dots,d_A \}$ that includes all $\omega \in 
\mathcal{P}(d_A)$ involving $i$.  
Thus,  $O_i\subset  \mathcal{P} \left( d_A\right)$ satisfies the relation $\omega \in O_i $ if and only if 
$ i \in \omega$ for $\omega \in \mathcal{P}(d_A)$. 
$\xi_c(p\|q)$ and $A_c(r|p\|q)$ are the classical Chernoff
\cite{C52} and Hoeffding bounds \cite{B74,CL71} given as 
\begin{align}
 \xi_c(p\|q) &= \sup_{0 \le s \le 1} -\phi(s|p\|q), \nonumber \\
 A_c(r|p\|q) &= \sup_{0 \le s \le 1} \frac{-sr-\phi(s|p\|q)}{1-s}, 
 \label{eq def xi c A c}
\end{align}
where $\phi(s|p\|q)$ is defined as $\phi(s|p\|q)=\log \sum_{\delta \in 
\Delta}p(\delta)^{1-s}q(\delta)^s$, $\Delta$ is a finite probability space of 
$p$ and $q$, and $\delta \in \Delta$ is an element of probability space.

The optimal error exponents $\tilde{\xi}_{\leftrightarrow}\left(\rho_{mix}\| \Psi \right)$ and 
 $\tilde{A}_{\leftrightarrow}\left(r | \rho_{mix} \| \Psi \right)$ are 
 nothing more than the Chernoff and Hoeffding bounds for
the particular class of two-way LOCC POVMs defined above, respectively.
Moreover, by their definitions, they are upper and lower bounded by
the optimal error exponents of one-way and two-way LOCC POVM, respectively: 
\begin{Lemma}\label{corollary two way LOCC}
 \begin{align}
\xi_{\rightarrow}  \left(\rho_{mix} \| \Psi \right) \le & 
 \tilde{\xi}_{\leftrightarrow} \left(\rho_{mix} \| \Psi \right)  \le
  \xi_{\leftrightarrow}\left(\rho_{mix} \| \Psi \right), \label{eq two way tilde xi}\\
A_{\rightarrow}\left(r|\rho_{mix} \| \Psi \right) \le & \tilde{A}_{\leftrightarrow}\left(r|\rho_{mix} \| \Psi \right) \le
 A_{\leftrightarrow}\left(r|\rho_{mix} \| \Psi \right). \label{eq two way tilde A}
 \end{align}
 In the above formulas, $\xi_{C}(\rho\|\sigma)$,  
$A_{C}(r|\rho\|\sigma)$, $\tilde{\xi}_{\leftrightarrow}(\rho_{mix}\|\Psi)$,and $\tilde{A}_{\leftrightarrow}(r|\rho_{mix}\|\Psi)$  
 are defined by Eqs.~(\ref{eq def xi C}), (\ref{eq def a c}), (\ref{eq 
 def tilde xi}), and (\ref{eq def tilde A}), respectively.
\end{Lemma}
{\bf (Proof)} \\
 The proofs of Eqs.~(\ref{eq two way tilde xi}) and (\ref{eq two way
 tilde A}) are essentially the same. Their second inequalities come
 directly from their definitions,
  since   $\tilde{\xi}_{\leftrightarrow}(\rho_{mix}\|\Psi)$ and
 $\tilde{A}_{\leftrightarrow}(r|\rho_{mix}\|\Psi)$ are optimal Chernoff and Hoeffding
 error exponents for the particular class of two-way LOCC POVMs.  

On the other hand, Lemma \ref{sec one-way lemma} and the discussion 
after Lemma \ref{sec one-way lemma} guarantee that 
the hypothesis testing under one-way LOCC POVMs between $\rho_{mix}$ and 
$\ket{\Psi}$ is equivalent to the classical hypothesis testing between 
the probability distributions $p_{mix}(i,j)$ and $q_\Psi(i,j)$. 
Here, $p_{mix}(i,j)$ and $q_\Psi(i,j)$ are defined as 
$p_{mix}(i,j)=\frac{1}{d_Ad_B}$ and $q_\Psi(i,j)=\lambda_i \delta_{i,j}$, 
respectively, where $\delta_{i,j}$ is the Kronecker delta, and the probability space consists of  all $(i,j)$ 
satisfying $1\le i \le d_A$ and $1 \le j \le d_B$. 
Actually, we can easily see that this classical hypothesis testing can 
be further reduced  
to the classical hypothesis between $p'_{mix}(k)$ and $q'_\Psi(k)$. 
Here, an element of probability space $k$ satisfies $1\le k \le 
R_s(\ket{\Psi})+1$, and  $p'_{mix}(k)$ and $q'_\Psi(k)$ are defined as  
\begin{align} 
p'_{mix}(k) & =\frac{1}{d_Ad_B} \quad \left (1\le \forall k \le 
 R_s(\ket{\Psi})\right ) \nonumber \\
p'_{mix}(R_s(\ket{\Psi})+1)& =\frac{d_Ad_B-d}{d_Ad_B} \nonumber \\
q'_\Psi(k) &=\lambda _k \quad \left (1\le \forall k \le 
 R_s(\ket{\Psi})\right ) \nonumber \\
q'_\Psi(R_s(\ket{\Psi})+1) &=0. \label{eq def p' mix q' psi}
\end{align}
Now, suppose Alice and Bob implement the three-step LOCC 
protocol by choosing the parameters 
${\bf m}=\{m_\omega(i)\}_{\omega,i}$ as 
\begin{align}
 m_\omega(i)=1,& \quad \mbox{if }\omega = \{ 1 \}, \dots, \{ d_A \}, \nonumber \\
m_\omega(i)=0,& \quad \mbox{otherwise}. \label{eq def optim one way}
\end{align}
Then, the classical probability distributions of the measurement 
outcomes are essentially described by Eq.~(\ref{eq def p' mix q' psi}). 
Thus, error exponents 
under the above choice of parameters ${\bf m}=\{m_\omega(i)\}_{\omega,i}$ 
equals the optimal error exponents for one-way LOCC POVMs.
Hence, the first inequalities of Eqs.~(\ref{eq two way tilde xi}) and (\ref{eq two way
 tilde A}) hold. 
\hfill $\square$

\subsection{Characterization of our optimal protocol}
By substituting Eqs.~(\ref{eq def P delta Psi m omega}) and (\ref{eq def P delta rho_mix m omega})
into Eqs.~(\ref{eq def tilde xi}) and (\ref{eq def tilde A}), and using Eq.~(\ref{eq def xi c A c}),
the bounds
$\tilde{\xi}_{\leftrightarrow}(\rho_{mix}\|\Psi)$
and
$\tilde{A}_{\leftrightarrow}(r|\rho_{mix}\|\Psi)$
are simplified as follows:
\begin{align}
&\tilde{\xi}_{\leftrightarrow}(\rho_{mix}\|\Psi) \nonumber \\
 = & \sup_{0 \le s \le 1, {\bf m}}
 \Big  \{f(s,{\bf m})\Big | 0 \le m_\omega(i) \le 1,
 \sum_{\omega \in O_i} m_\omega(i) \le 1 \Big \} \nonumber \\
 = & \sup_{0 \le s \le 1}f(s), \label{eq tilde xi = sup s}  \\
&\tilde{A}_{\leftrightarrow}(r|\rho_{mix}\|\Psi) \nonumber \\
 = &\sup_{0 \le s \le 1, {\bf m}}
 \Big  \{\frac{f(s,{\bf m})-rs}{1-s} \Big |
 0 \le m_\omega(i) \le 1,
 \sum_{\omega \in O_i} m_\omega(i) \le 1 \Big \}  \nonumber \\
 = &\sup_{0 \le s \le 1}
\frac{f(s)-rs}{1-s},
  \label{eq tilde A = sup s}
\end{align}
where objective functions $f(s,{\bf m})$ and $f(s)$
are given as
\begin{align}
 & f(s,{\bf m}) \nonumber \\
:=&- \log \Tr P(\delta|\rho_{mix}, {\bf m})^{1-s} P(\delta|\Psi, {\bf m})^s \nonumber \\
=&- \log \Big[ 
(d_Ad_B)^{s-1}\cdot \sum_{\omega \in
 \mathcal{P}(d_A)}|\omega|^{1-s} \nonumber\\
& \qquad \cdot
\Big( \sum_{i \in \omega} (m_\omega(i))^2 \cdot \lambda_i \Big)^{1-s}\cdot \Big ( \sum_{i \in \omega} m_\omega(i)\cdot \lambda_i
 \Big)^{2s-1} \Big], \label{eq two way f s omega = -log} \\
 & f(s) \nonumber \\
:=&
 \max_{ {\bf m}}
 \Big  \{f(s,{\bf m})\Big | 0 \le m_\omega(i) \le 1,
 \sum_{\omega \in O_i} m_\omega(i) \le 1 \Big \}. \label{eq two way f s  = max}
\end{align}
We also define 
\begin{align}
{\bf m}_s:= 
 \argmax_{ {\bf m}}
 \Big  \{f(s,{\bf m})\Big | 0 \le m_\omega(i) \le 1,
 \sum_{\omega \in O_i} m_\omega(i) \le 1 \Big \}.
\end{align}
Note that the uniqueness of ${\bf m}_s$ does not hold in general.
However, for $s \in [\frac{1}{2},1]$, 
we can show the following lemma.
\begin{Lemma}\label{lem9}
For $s \in [\frac{1}{2},1]$, we have
\begin{align}
f(s,{\bf m})\le - \log \sum_i \lambda_i^s +(1-s)\log (d_A d_B).
\end{align}
The equality holds for $s \in [\frac{1}{2},1)$
only when 
the measurement parameter ${\bf m}$ satisfies condition 
(\ref{eq def optim one way}). 
\end{Lemma}

{\bf (Proof)} \\
Using the H\"{o}lder inequality, we have 
\begin{align*}
&|\omega|^{1-s} \cdot
\Big( \sum_{i \in \omega} (m_\omega(i))^2 \cdot \lambda_i \Big)^{1-s}\cdot \Big ( \sum_{i \in \omega} m_\omega(i)\cdot \lambda_i
 \Big)^{2s-1} \\
=&|\omega|^{1-s} \cdot
\Big( \sum_{i \in \omega} (m_\omega(i))^2 \cdot \lambda_i \Big)^{s \cdot \frac{1-s}{s}}\cdot \Big ( \sum_{i \in \omega} m_\omega(i)\cdot \lambda_i
 \Big)^{s \cdot \frac{2s-1}{s}} \\
=&|\omega|^{1-s} \cdot
\Big( \sum_{i \in \omega} (m_\omega(i))^2 \cdot \lambda_i \Big)^{s \cdot \frac{1-s}{s}}\cdot \Big ( \sum_{i \in \omega} m_\omega(i)\cdot \lambda_i
 \Big)^{s \cdot \frac{2s-1}{s}} \\
\ge &|\omega|^{1-s} \cdot
\Big( \sum_{i \in \omega} (m_\omega(i))^{2\frac{1-s}{s}} \cdot 
\lambda_i^{\frac{1-s}{s}} 
\cdot m_\omega(i)^{\frac{2s-1}{s}}\cdot \lambda_i^{\frac{2s-1}{s}}
 \Big)^{s} \\
= &(\sum_{i \in \omega}1)^{1-s} \cdot
\Big( \sum_{i \in \omega} (m_\omega(i))^{\frac{1}{s}} \cdot 
\lambda_i \Big)^{s} \\
\ge &
\sum_{i \in \omega}1^s (m_\omega(i))^{\frac{s}{s}} \cdot 
\lambda_i^s 
= 
\sum_{i \in \omega} m_\omega(i) \cdot 
\lambda_i^s . 
\end{align*}
Thus, 
\begin{align*}
& e^{-f(s,{\bf m})}
\ge (d_Ad_B)^{s-1}\cdot \sum_{\omega \in
 \mathcal{P}(d_A)} \sum_{i \in \omega} m_\omega(i) \cdot 
\lambda_i^s  \\
=& (d_Ad_B)^{s-1}\cdot \sum_{i =1}^{d_A} \lambda_i^s .
\end{align*} 
The equality condition follows from the equality condition for the H\"{o}lder inequality.
\hfill $\square$

Lemma \ref{lem9} guarantees that
$f(s)=- \log \sum_i \lambda_i^s +(1-s)\log (d_A d_B)$ 
for $s \in [\frac{1}{2},1]$.
For $s \in [\frac{1}{2},1)$,
${\bf m}_s$ is uniquely determined to be the measurement parameter ${\bf 
m}$ satisfying condition  
(\ref{eq def optim one way}). 
In this case, 
${\bf m}_s$ does not depend on $s$ and 
is described as  ${\bf m}_{\to}$
because it can be realized by a one-way LOCC.
Now, we consider our problem 
by using the important parameter 
\begin{equation}\label{eq two way sr argmax}
s_r:=\argmax_{0\le s\le 1}\frac{f(s)-rs}{1-s}.
\end{equation}
When $s_r$ belongs to $[\frac{1}{2},1]$,
the optimal Hoeffding bound can be attained by 
a one-way LOCC POVM, i..e, 
the equality in the first inequality in (\ref{eq two way tilde A}) holds.

Next, as another typical case, we consider the measurement corresponding to 
${\bf m}_0$.
Since 
\begin{align}
e^{-f(0,{\bf m})}
= 
\sum_{\omega \in \mathcal{P}(d_A)}
\frac{|\omega|  \sum_{i \in \omega} (m_\omega(i))^2 \cdot \lambda_i }
{d_A d_B \sum_{i \in \omega} m_\omega(i)\cdot \lambda_i}
\end{align}
is the type-1 error probability (the error when the true is $\rho_{mix}$)
under the constraint that the type-2 error is zero,
${\bf m}_0$ corresponds to the measurement
minimizing the type-1 error probability 
under the constraint that the type-2 error is zero.
That is, 
\begin{align}
{\bf m}_0=
\argmin_{{\bf m}}
\sum_{\omega \in \mathcal{P}(d_A)}
\frac{|\omega|  \sum_{i \in \omega} (m_\omega(i))^2 \cdot \lambda_i }
{d_A d_B \sum_{i \in \omega} m_\omega(i)\cdot \lambda_i},
\label{m0}
\end{align}
which is essentially discussed in the previous paper \cite{OH11}.
That is,
${\bf m}_s$ gives a new two-way LOCC measurement only for $s \in (0, \frac{1}{2})$.
Therefore, we can consider the following three distinct cases:
\begin{description}
\item[Case 1:]
$s_r$ belongs to $[\frac{1}{2},1]$.

\item[Case 2:]
$s_r$ belongs to $(0, \frac{1}{2})$.

\item[Case 3:]
$s_r$ is $0$.
\end{description}
Only in Case 2, the optimal parameter is different from 
${\bf m}_0$ and ${\bf m}_{\to}$.
That is, only in Case 2, our two-way LOCC measurement
has a gain over one-way LOCC and the two-way LOCC POVMs given in the previous paper \cite{OH11}.
To characterize Case 2,
we introduce the notations.
We denote the first-step POVM defined by
${\bf m}_s$, ${\bf m}_0$, and ${\bf m}_{\to}$
via (\ref{eq def M omega})
by $M(s)$, $M(0)$, and $M(\to)$, respectively.
For a POVM $M=\{M_{\omega}\}$, 
the Choi-Jamiolkowski matrix of the instrument is defined by the POVM $M$; 
that is,  
\begin{align}
 J(M)
\stackrel{\rm def}{=}& \sum_{\omega \in 
  \mathcal{P}(d_A)}\ket{\omega}_M\bra{\omega}_M\otimes 
 \left(\sqrt{M_\omega} \otimes I_R \right) \nonumber \\
& \quad \cdot \ket{\Phi^+_{d_A}}_{AR}\bra{\Phi^+_{d_A}}_{AR} \cdot \left(\sqrt{M_\omega} \otimes I_R \right),
\end{align}
where $\{\ket{\omega}_M\}_{\omega \in \mathcal{P}(d_A)}$ is the 
orthonormal basis of the system of the measurement device $\Hi_M$,
$I_R$  is the identity of the reference system $\Hi_R$, and $\ket{\Phi^+_{d_A}}_{AR}$
is a maximally entangled state on the joint system of Alice's system
and a reference system $\Hi_A\otimes \Hi_R$. 
To see the differences of the optimal POVM 
$M(s)$ from $M(0)$ and $M(\to)$,
we define the Jamiolkowski distances \cite{GLN05}: 
\begin{align}
D_1(r) &\stackrel{\rm def}{=}\frac{1}{2}\|J(M_{s_r})-J(M_0)\|_1, \\
D_2(r) &\stackrel{\rm def}{=}\frac{1}{2}\|J(M_{s_r})-J(M_{\to})\|_1,
\end{align} 
where $\|\cdot \|_1$ is the trace norm. 
The parameters $D_1(r)$ and $D_2(r)$
describe how different from the one-way case and the case (\ref{m0})
the optimal measurement is.
These are numerically evaluated in our concrete example in later.

We characterize $s_r$ in Case 1.
Taking the derivative, we have
\begin{align}
\frac{d}{ds}(\frac{f(s)-sr}{1-s})
=\frac{f'(s)(1-s) -r +f(s)}{(1-s)^2}.
\end{align}
Since $\frac{d}{ds}(f'(s)(1-s)+f(s))=f''(s)(1-s)<0$,
when the derivative 
$\frac{d}{ds}(\frac{f(s)-sr}{1-s})$ is zero at $s$,
$s$ realizes the maximum of $\frac{f(s)-sr}{1-s}$.
That is,  
$s_r$ satisfies
\begin{align}
r=f'(s_r)(1-s_r)+f(s_r).
\label{eq3-22-1}
\end{align}
Then, the right hand side of (\ref{eq3-22-1}) is strictly and monotonically decreasing.
Hence, $s_r$ is  strictly and monotonically decreasing.
In particular, 
since $f'(1)(1-1)+f(1)=0$, $\lim_{r \to 0}s_r=1$.

Now, we consider the situation when Case 2 vanishes.
Note that 
the function $f(s)$ is not necessarily concave in $s\in (0,\frac{1}{2})$,
while it is concave in $s\in (\frac{1}{2},1)$.
Since
\begin{align*}
& \frac{d}{dr}\max_{\frac{1}{2}\le s \le 1} \frac{f(s)-sr}{1-s}
=\frac{d}{dr}\frac{f(s_r)-s_rr}{1-s_r} \\
=&
\frac{\partial }{\partial r}\frac{f(s)-sr}{1-s}|_{s=s_r}
+
\frac{d s_r}{dr}
\frac{\partial }{\partial s}\frac{f(s)-sr}{1-s}|_{s=s_r}\\
=&
\frac{\partial }{\partial r}\frac{f(s)-sr}{1-s}|_{s=s_r}
=
\frac{-s_r}{1-s_r},
\end{align*}
the value 
$\max_{\frac{1}{2}\le s \le 1} \frac{f(s)-sr}{1-s}$
decreases more rapidly 
than $f(0)-r$.
When $
\max_{\frac{1}{2}\le s \le 1} \frac{f(s)}{1-s}=
f'(1)= E(|\Psi\rangle)+ \log (d_A d_B) \le f(0)$,
we have
$\max_{\frac{1}{2}\le s \le 1} \frac{f(s)-s r}{1-s}
\le f(0)-r$.
That is, Case 1 does not appear.
When $f(0)$ 
is smaller than $f'(1)$ but is sufficiently large,
there is a point $r_*$ such that
$\max_{\frac{1}{2}\le s \le 1} \frac{f(s)-s r_*}{1-s}
=f(0)-r_*$.
In this case,
when $r < r_*$,
$s_r$ belongs to Case 1.
When $r \ge r_*$,
$s_r$ belongs to Case 3.
That is, the parameter $s_r$ suddenly goes to zero at $r=r_*$.
However, when Case 2 does not vanish,
our situation is more complicated.

\subsection{Numerical evaluation}
In this final part of this section, we present the results of numerical 
calculations of $s$ and ${\bf m}$ 
achieving $\tilde{\xi}_{\leftrightarrow}(\rho_{mix}\|\Psi)$ and 
$\tilde{A}_{\leftrightarrow}(r|\rho_{mix}\|\Psi)$. 
Note that the plot of $\tilde{\xi}_{\leftrightarrow}(\rho_{mix}\|\Psi)$ and 
$\tilde{A}_{\leftrightarrow}(r|\rho_{mix}\|\Psi)$ are given in the next 
section with the plot of all other error exponents.

Numerical calculations have been performed for the following one-parameter family of 
pure states $\ket{\Psi(\lambda)}$ up to $d_A=d_B=4$:
\begin{equation}\label{eq def Psi lambda}
 \ket{\Psi(\lambda)}=\sqrt{\lambda}\left(\sum_{i=1}^{d-1} \ket{ii} \right)+\sqrt{1-\lambda}\ket{dd},
\end{equation}
where $\lambda$ satisfies $0\le \lambda \le 1/\sqrt{d}$ and $d=\min 
\left( d_A, d_B \right)$. Note that $\ket{\Psi(0)}$ is a product state
and $\ket{\Psi(1/\sqrt{d})}$ is a maximally entangled state. 

We first calculate the optimal $s$ achieving $\tilde{\xi}_{\leftrightarrow}(\rho_{mix}\|\Psi(\lambda))$
in terms of Eq.~(\ref{eq tilde xi = sup s}).
Then, we observed that $s=0$ attains the optimum for all $\lambda$ in 
low-dimensional systems up to $d_A=d_B=4$. 
That is, the optimal POVM is given by ${\bf m}_0$, 
which is characterized by (\ref{m0}).

\begin{figure}[htbp]
\centering
\includegraphics[width=\linewidth]{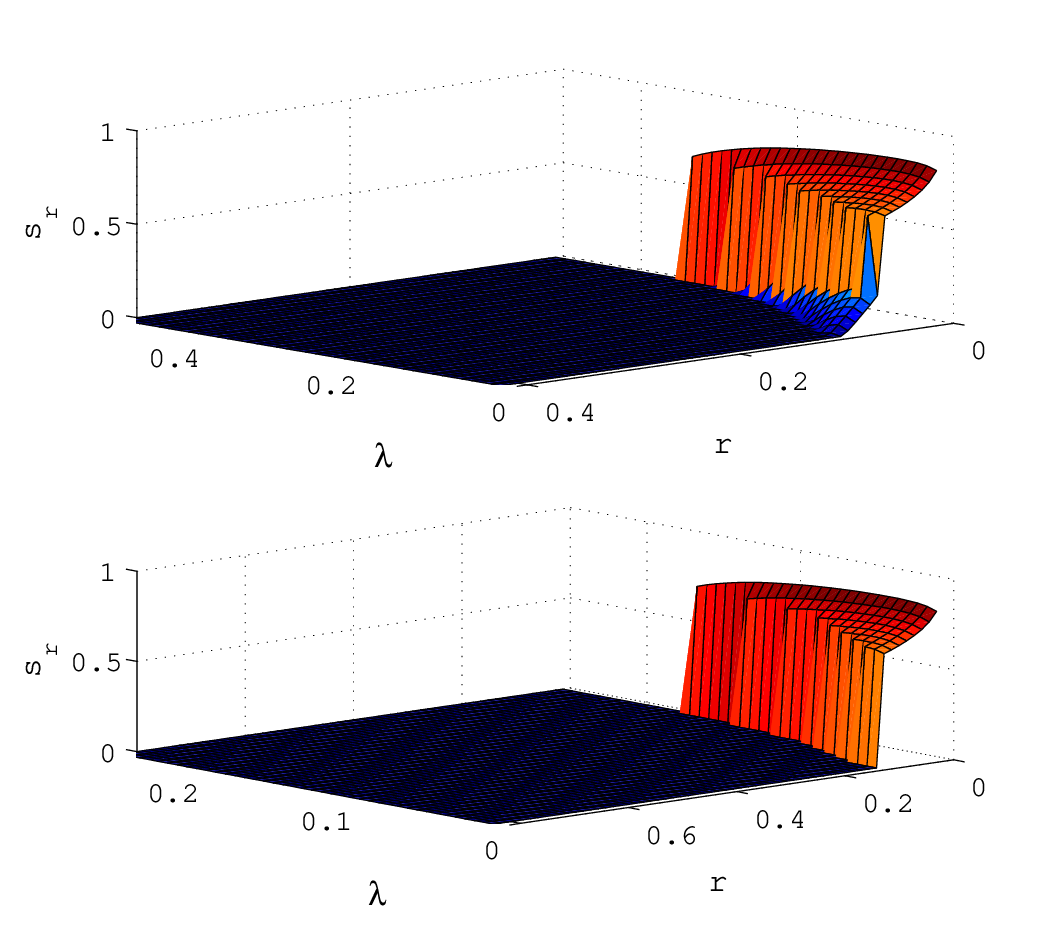}
\caption{(Color online) Plots of the optimal $s$ achieving 
the Hoeffding bound $\tilde{A}_{\leftrightarrow}\left (r|\rho_{mix} \| \Psi(\lambda)\right)$ with respect to $r$ 
 and $\lambda$, where $\ket{\Psi(\lambda)}$ is defined by Eq.~(\ref{eq def Psi lambda}).
The upper panel is the plot when $d_A=d_B=2$, and the lower 
one is the plot when $d_A=d_B=4$.}
\label{fig optimal s}
\end{figure}

Next, we calculate 
the parameter $s_r$, i.e.,  
the optimal $s$ achieving 
the Hoeffding bound 
$\tilde{A}_{\leftrightarrow}(r|\rho_{mix}\|\Psi(\lambda))$
as a function of $r$ and $\lambda$
for $d=d_A=d_B=2$ and $d=d_A=d_B=4$ in 
FIG. \ref{fig optimal s}.
As a result,
Case 2 appears for $d=d_A=d_B=2$ in FIG. \ref{fig optimal s}
but it does not appear for $d=d_A=d_B=4$.
This means that 
our new measurement strategy is needed in order to attain 
the Hoeffding bound 
$\tilde{A}_{\leftrightarrow}(r|\rho_{mix}\|\Psi(\lambda))$
at least for $d=d_A=d_B=2$.

\begin{figure}[htbp]
\centering
\includegraphics[width=\linewidth]{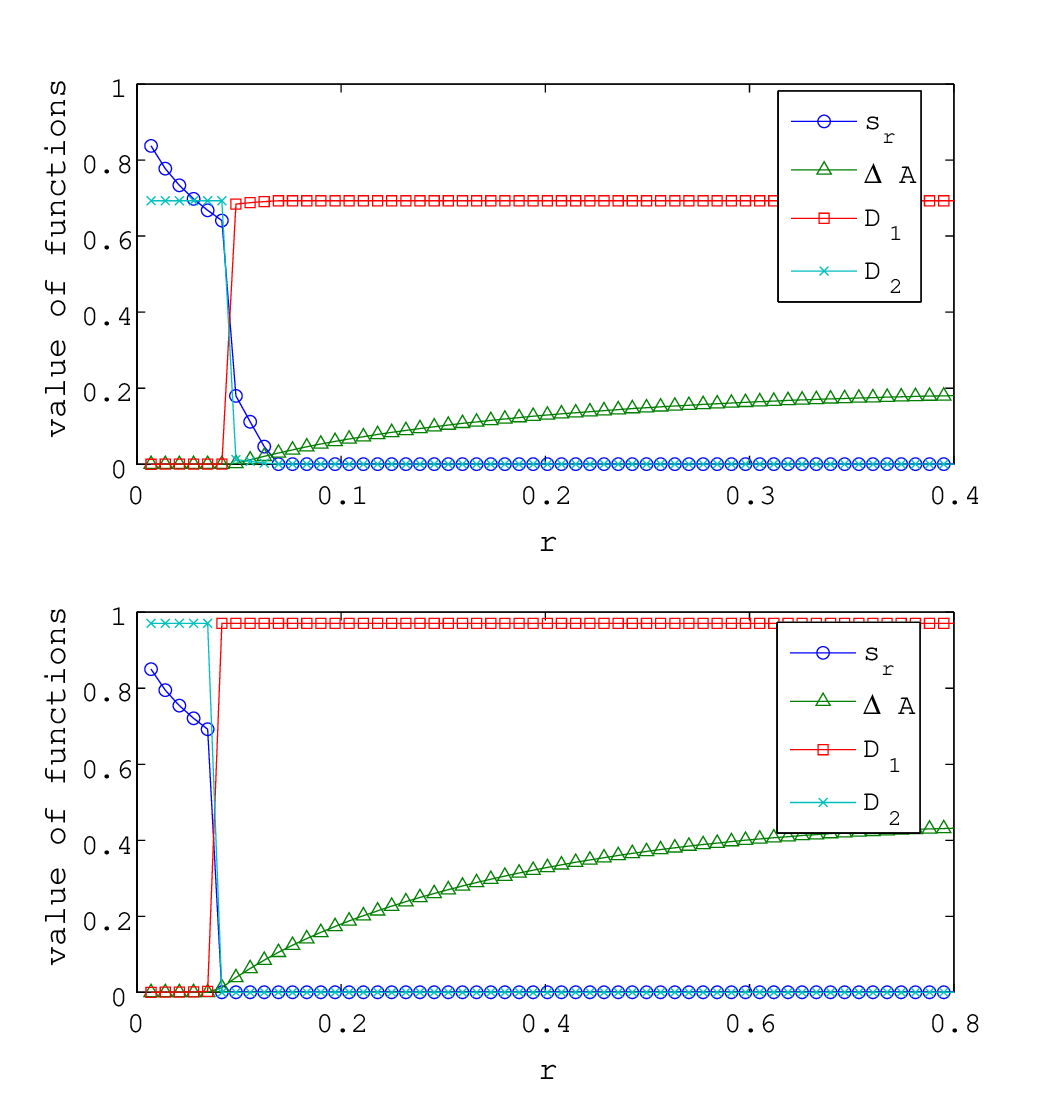}
\caption{(Color online) Plots of $s_r$,$\Delta A(r)$, 
$D_1(r)$, $D_2(r)$ with respect to $r$ for 
$d=d_A=d_B=2$ (upper panel) and $d=d_A=d_B=4$. 
In the upper panel, $\lambda$ is chosen as $\lambda=0.1$.
% and $r_2$, which appears in the definition of $D_2(r,\lambda)$, is chosen as  $r_2=0.4$. 
Similarly, in the lower one, $\lambda$ is chosen as $\lambda=0.05$.}
\label{fig optimal povm}
\end{figure}

To see the relation between $r$ and $s_r$, 
we calculate $s_r$ with the fixed $\lambda=0.1$. As shown in FIG. 
\ref{fig optimal povm}, the parameter $s_r$ gradually decreases when $r$ 
increases, and at some value of $r$, $s_r$ suddenly goes down to $0$
for $d=d_A=d_B=4$.
However,
for $d=d_A=d_B=2$,
though the parameter $s_r$ is gradually decreasing when $r$ 
increases, $s_r$ slowly approaches $0$.


In order to see the detailed clear separation of 
our two-way LOCC POVM performance from our one-way LOCC POVM,
we perform further numerical calculations with a fixed $\lambda=0.1$
(FIG \ref{fig optimal povm}).
We numerically calculate the difference between $\tilde{A}_{\leftrightarrow}(r|\rho_{mix}\|\Psi(\lambda))$
and $A_{\rightarrow}(r|\rho_{mix}\|\Psi(\lambda))$:
\begin{equation}
 \Delta A(r)\stackrel{\rm 
  def}{=}\tilde{A}_{\leftrightarrow}(r|\rho_{mix}\|\Psi(\lambda))-
A_{\rightarrow}(r|\rho_{mix}\|\Psi(\lambda)) 
\end{equation}
To estimate the difference in the optimal measurement from 
the measurements $M(0)$ and $M(\to)$,
we also numerically calculate 
$D_1(r)$ and $D_2(r)$.

FIG.~\ref{fig optimal povm} shows plots of $\Delta A(r)$, 
$D_1(r)$, $D_2(r)$ as well as of $s_r$ with respect to $r$ for
$d=d_A=d_B=2$ (upper panel) and $d=d_A=d_B=4$ (lower panel). 
Further, in FIG.~\ref{fig optimal povm}, $\lambda$ is fixed as 
$\lambda=0.1$ for $d=2$, and as  
$\lambda=0.05$ for $d=4$. 
Note that we choose very smaall $\lambda$ in FIG.~\ref{fig optimal povm}
because FIG.~\ref{fig optimal 
s} suggests that Case 1 never appears when $\lambda$ is large.   

From FIG.~\ref{fig optimal povm}, 
we observe that there are three different regions (Cases 1, 2, and 3)
when $d=d_A=d_B=2$: 
When $r$ increases from $r=0$,
firstly, we find the Case-1 region where $s_r$ is far from $0$ and decreases slowly, 
and  $\Delta A(r)=D_1(r)=0$.
That is, we can numerically confirm that 
there is no benefit from using 
two-way LOCC POVM as long as it is in the class of the three-step LOCC 
protocols introduced in this section. 
Secondly, we find the Case-2 region where 
$s_r$ and $D_2(r)$ take small but non-zero values, 
$D_1(r)$ is far from $0$, and $\Delta A(r)\neq 0$.
Finally, we find the Case-3 region where 
$s_r=D_2(r)=0$, but $\Delta A(r)\neq 0$.
Because $D_2(r)=0$, we can numerically confirm that
the optimal POVM does not depend on $r$ in this region.

In FIG.\ref{fig optimal povm}, we can observe the Case-2 region when $d=2$ 
but we can hardly observe the intermediate region when $d=4$. 
In other words, in the lower panel of FIG.\ref{fig optimal povm}, 
around $r=0.12$, $s_r$ suddenly jumps from the 
values of more than $0.5$ to $0$, and at the same time, 
$D_1(r)$ suddenly jumps from $0$ to $1$, and
$D_2(r)$ suddenly jumps from $1$ to $0$.
From this observation, we can conjecture that 
Case 2 does not appear when dimensions $d_A$ and $d_B$ are large.
In other words, for a fixed $\lambda$, 
we can essentially attain the Hoeffding bound $\tilde{A}_{\leftrightarrow}(r|\rho_{mix}\|\Psi(\lambda))$ 
using only two POVMs:
The one-way LOCC POVM defined by Eq.~(\ref{eq def optim one way}),
and the two-way LOCC POVM attaining the optimum of 
Eq.~(\ref{m0}).
The former attains the optimum of $\tilde{A}_{\leftrightarrow}(r|\rho_{mix}\|\Psi(\lambda))$ 
in the Case-1 region (with small $r$) and the latter attains the optimism in the Case-3 
region (with large $r$).

\section{Plots of the error  exponents} \label{section plot}
In this section, we give plots of various error exponents for our local 
hypothesis testing problem. 
Since there is no difference among one-way LOCC, 
two-way LOCC, and separable POVMs in terms of the error exponents corresponding to Stein's Lemma,
here we only plot error exponents corresponding to the Chernoff and 
Hoeffding bounds.

\begin{figure}[htbp]
\centering
\includegraphics[width=\linewidth]{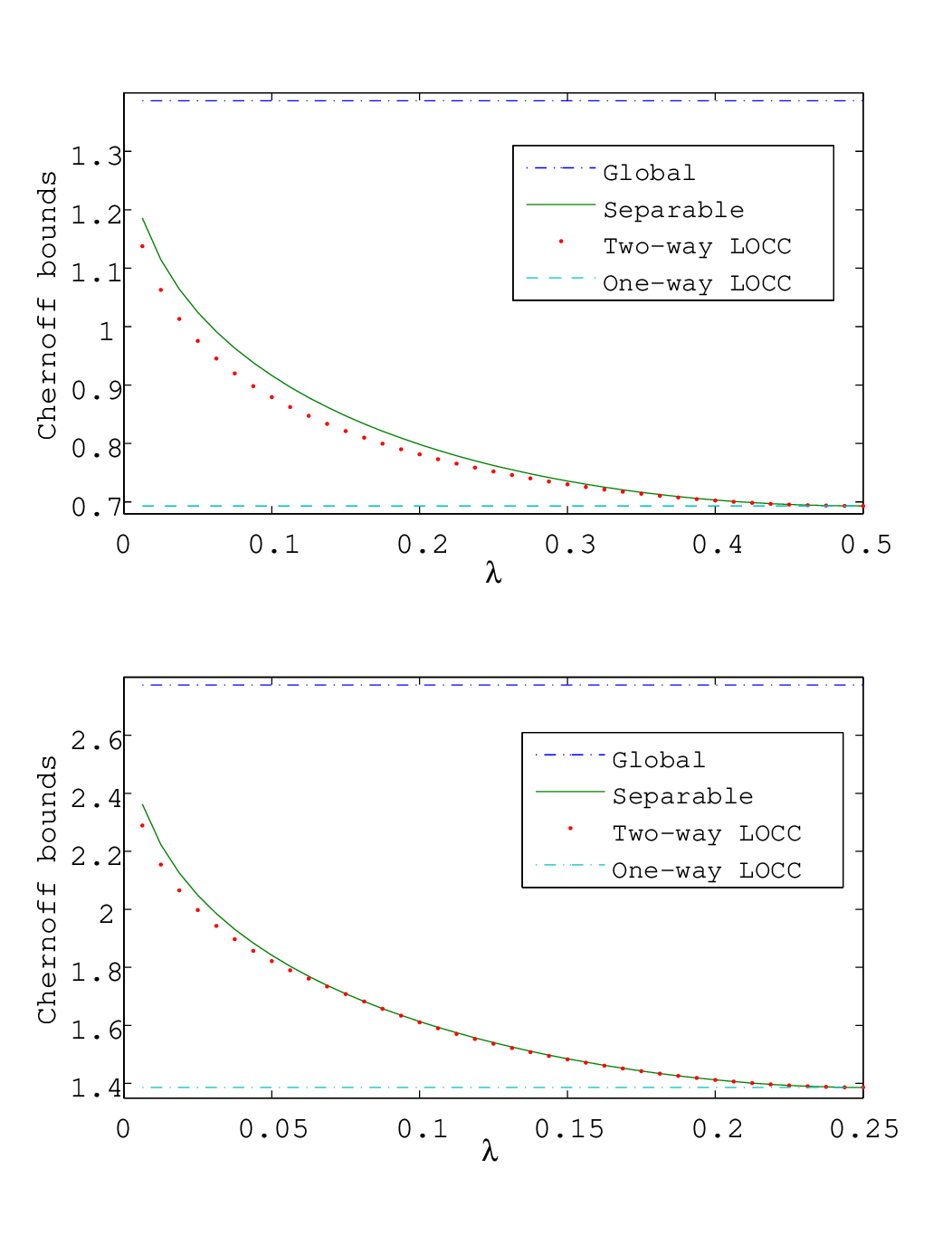}
\caption{(Color online) Chernoff bounds for $\ket{\Psi(\lambda)}$ 
 defined  by Eq.~(\ref{eq def Psi lambda})) for $d_A=d_B=2$ (upper panel)
and $d_A=d_B=4$ (lower panel). 
The lines labeled ``Global'', ``Separable'', ``Two-way LOCC'', and
 ``One-way LOCC'' are plots of $\xi_g(\rho_{mix}\|\Psi(\lambda))$, $\xi_{sep}(\rho_{mix}\|\Psi(\lambda))$,
 $\tilde{\xi}_{\leftrightarrow}(\rho_{mix}\|\Psi(\lambda))$, 
 $\xi_{\rightarrow}(\rho_{mix}\|\Psi(\lambda))$ as functions of $\lambda$, 
 respectively. } 
\label{fig cher}
\end{figure}

\begin{figure}[htbp]
\centering
\includegraphics[width=\linewidth]{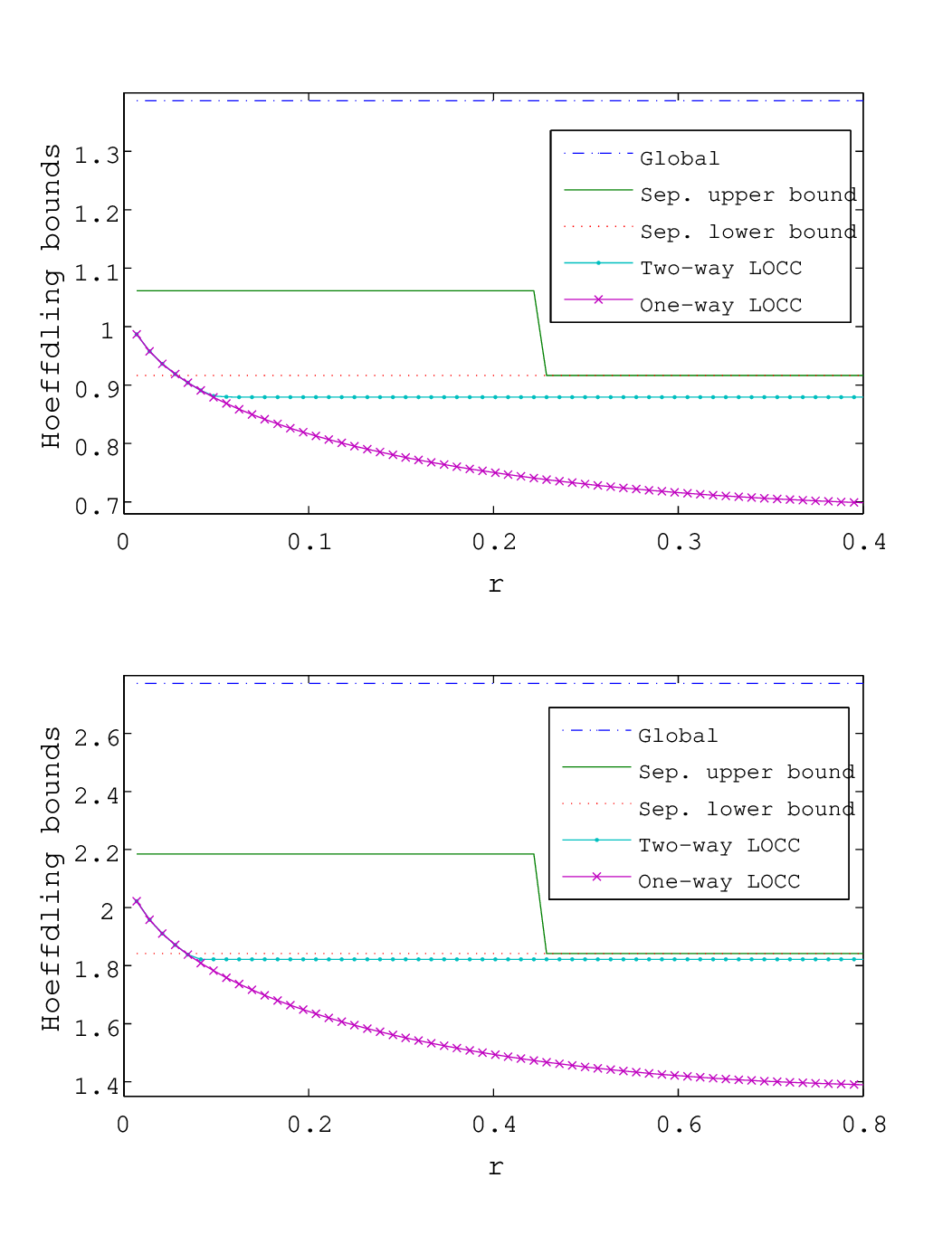}
\caption{(Color online) Hoeffding bounds for $\ket{\Psi(\lambda )}$. 
 The upper panel is for $d_A=d_B=2$ and $\lambda=0.1$,
and the lower one is for $d_A=d_B=4$ and $\lambda=0.05$. 
The lines labeled ``Global'', ``Sep. upper bound'',``Sep. lower bound'', ``Two-way LOCC'', and
 ``One-way LOCC'' are plots of $A_g(r|\rho_{mix}\|\Psi(\lambda))$, 
 $\overline{A}_{sep}(r|\rho_{mix}\|\Psi(\lambda))$, $\underline{A}_{sep}(r|\rho_{mix}\|\Psi(\lambda))$, 
 $\tilde{A}_{\leftrightarrow}(r|\rho_{mix}\|\Psi(\lambda))$, $A_{\rightarrow}(r|\rho_{mix}\|\Psi(\lambda))$, respectively.}
\label{fig hoeff}
\end{figure}

FIG.~\ref{fig cher} shows plots of the Chernoff bounds 
$\xi_g\left(\rho_{mix}\|\Psi(\lambda) \right)$, $\xi_{sep}\left(\rho_{mix}\|\Psi(\lambda) \right)$,
 $\tilde{\xi}_{\leftrightarrow}\left(\rho_{mix}\|\Psi(\lambda) \right)$, 
 $\xi_{\rightarrow}\left(\rho_{mix}\|\Psi(\lambda) \right)$ as functions of 
 $\lambda$ when $d=d_A=d_B=2$ (upper panel) and $d=d_A=d_B=4$ (
 lower panel), where $\xi_C\left(\rho \|\sigma \right)$, 
 $\tilde{\xi}_{\leftrightarrow}\left(\rho_{mix}\|\Psi \right)$, 
 and $\Psi(\lambda)\stackrel{\rm  
 def}{=}\ket{\Psi(\lambda)}\bra{\Psi(\lambda)}$ are defined by 
 Eqs.~(\ref{eq def xi C}), (\ref{eq def tilde xi}) and (\ref{eq def Psi 
 lambda}), respectively. 
$\xi_{\rightarrow}\left(\rho_{mix}\|\Psi \right)$ and  
$\xi_{sep}\left(\rho_{mix}\|\Psi \right)$ are calculated via Eqs.~(\ref{eq xi rightarrow}) and (\ref{eq xi sep = log d A log  
dB -LR}), respectively.    
On the other hand, $\tilde{\xi}_{\leftrightarrow}\left(\rho_{mix}\|\Psi \right)$ is numerically calculated by 
means of  Eqs.~(\ref{eq tilde xi = sup s})
and (\ref{eq two way f s omega = -log}).
Here, we observe that $\tilde{\xi}_{\leftrightarrow}\left(\rho_{mix}\|\Psi(\lambda) \right)$ is
 always strictly larger than $\xi_{\rightarrow}\left(\rho_{mix}\|\Psi(\lambda) \right)$ except when
 $\ket{\Psi(\lambda)}$ is a product state ($\lambda=0$) or a maximally 
 entangled state ($\lambda=1/d$).
Moreover, $\tilde{\xi}_{\leftrightarrow}\left(\rho_{mix}\|\Psi \right)$ well
 approximates $\xi_{sep}\left(\rho_{mix}\|\Psi \right)$. This means that the three-step LOCC
 protocols defined in the previous section not only outperform the best 
 one-way LOCC protocols, but also 
 are even near the optimum over all two-way LOCC protocols from the
 viewpoint of the Chernoff bounds.

Next we give plots of the Hoeffding bounds $A_C\left(r|\rho_{mix}\|\Psi 
\right)$ defined by Eq.~(\ref{eq def a c}) against parameter $r$. 
Before showing the plots, we give two remarks. First, from the definition, the Hoeffding bound is equal to the 
 Stein's Lemma type of the error exponent in the limit of $r \rightarrow +0$. 
Thus, the y-intercepts of the plots for one-way LOCC $A_\rightarrow\left(r|\rho_{mix}\|\Psi \right)$, 
two-way LOCC $A_\leftrightarrow\left(r|\rho_{mix}\|\Psi \right)$, and separable POVM $A_{sep}\left(r|\rho_{mix}\|\Psi \right)$
are all equal to 
$\eta_{\rightarrow}\left(\epsilon|\rho_{mix}\|\Psi 
\right)=\eta_{\leftrightarrow}\left(\epsilon|\rho_{mix}\|\Psi 
\right)=\eta_{sep}\left(\epsilon|\rho_{mix}\|\Psi \right) =\log d_A +\log d_B - E(\ket{\Psi})$,
although, as we can observe from FIG. \ref{fig hoeff}, the convergence of $A_{\rightarrow}\left(r|\rho_{mix}\|\Psi \right)$ in the limit $r 
\rightarrow +0$ is very slow.  
Second, when a given $r_0$ satisfies 
$A_C\left (r_0|\rho_{mix}\|\Psi \right)=r_0$, the Hoeffding bound for $r_0$ is equal to the Chernoff 
bound: $A_C\left (r_0|\rho_{mix}\|\Psi \right)=\xi _C\left (\rho_{mix}\|\Psi \right)$. 
We can actually find such $r_0$ as an intersection of the plots of 
$y=A_C\left (r|\rho_{mix}\|\Psi \right)$ and $y=x$.  Thus, we can derive 
the value of $\xi_C\left (\rho_{mix}\|\Psi \right)$  from 
the plots of $A_C\left(r|\rho_{mix}\|\Psi \right)$.

For one-way LOCC POVMs, we have an analytical expression of 
$A_{\rightarrow}\left(r|\rho_{mix}\|\Psi \right)$ via Theorem \ref{theorem one way}. 
On the other hand, for separable POVMs, since we do not know an analytical formula for $A_{sep}\left(r|\rho_{mix}\|\Psi \right)$
available for the whole range of $r >0$, we use analytical upper
 and lower bounds $\overline{A}_{sep}\left(r|\rho_{mix}\|\Psi \right)$ 
 and $\underline{A}_{sep}\left(r|\rho_{mix}\|\Psi \right)$ of
$A_{sep}\left(r|\rho_{mix}\|\Psi \right)$ derived from Corollary \ref{corollary A sep r} instead of $A_{sep}\left(r|\rho_{mix}\|\Psi \right)$: 
\begin{align}
&\overline{A}_{sep}\left(r|\rho_{mix}\|\Psi \right) \nonumber \\
\stackrel{\rm def}{=}&
\left \{
\begin{array}{l}
\log d_A + \log d_B -E(\ket{\Psi}), \\
 \qquad {\rm if} \ 0 \le r \le \log d-LR(\ket{\Psi})
\\
\quad  \\
 \log d_A + \log d_B  - LR(\ket{\Psi}), \\ 
\qquad  {\rm if} \ r \ge  \log d-LR(\ket{\Psi})\\
\end{array}
\right . \label{eq def overline A sep}\\
& \underline{A}_{sep}\left(r|\rho_{mix}\|\Psi \right)\nonumber \\
\stackrel{\rm def}{=}& \log d_A + \log d_B - LR(\ket{\Psi})\label{eq def underline A sep}
\end{align}
FIG.~\ref{fig hoeff} shows plots of the Hoeffding bounds 
$A_g\left(r|\rho_{mix}\|\Psi(\lambda) \right)$, 
$\overline{A}_{sep}\left(r|\rho_{mix}\|\Psi(\lambda) \right) $, 
$\underline{A}_{sep}\left(r|\rho_{mix}\|\Psi(\lambda) \right)$, 
 $\tilde{A}_{\leftrightarrow}\left(r|\rho_{mix}\|\Psi(\lambda) \right)$, 
 and $A_{\rightarrow}\left(r|\rho_{mix}\|\Psi(\lambda) \right)$ as 
 functions of 
 parameter $r$,  where $\Psi(\lambda)\stackrel{\rm  
 def}{=}\ket{\Psi(\lambda)}\bra{\Psi(\lambda)}$ is defined by 
 Eq.~(\ref{eq def Psi 
 lambda}).
In FIG.~\ref{fig hoeff}, the upper panel shows the plots for 
$d=d_A=d_B=2$ and $\lambda=0.1$, and 
the lower one shows them for $d=d_A=d_B=4$ and $\lambda=0.05$.
Here, $\tilde{A}_{\leftrightarrow}\left(r|\rho_{mix}\|\Psi \right)$ is 
numerically calculated by means of  Eqs.~(\ref{eq tilde A = sup s}) 
and (\ref{eq two way f s omega = -log}).

From FIG. \ref{fig hoeff}, we observe that $\tilde{A}_{\leftrightarrow}\left(r|\rho_{mix}\|\Psi(\lambda) \right)$ well
 approximates $\underline{A}_{sep}\left(r|\rho_{mix}\|\Psi(\lambda) 
 \right)$ if $A_{\rightarrow}\left(r|\rho_{mix}\|\Psi(\lambda) \right) < 
 \underline{A}_{sep}\left(r|\rho_{mix}\|\Psi(\lambda) \right)$;
otherwise $\tilde{A}_{\leftrightarrow}\left(r|\rho_{mix}\|\Psi(\lambda) 
\right)= A_{\rightarrow}\left(r|\rho_{mix}\|\Psi(\lambda) \right)$. 
As a result, $\tilde{A}_{\leftrightarrow}\left(r|\rho_{mix}\|\Psi(\lambda) 
\right)$ well approximates $A_{sep}\left(r|\rho_{mix}\|\Psi(\lambda)  
\right)=\overline{A}_{sep}\left(r|\rho_{mix}\|\Psi(\lambda) 
\right)=\underline{A}_{sep}\left(r|\rho_{mix}\|\Psi(\lambda) \right)$ 
 in the region of $r \ge  \log d- LR(\ket{\Psi(\lambda)})$.
Thus, we observe that, at least in low-dimensional systems and in this
 region of parameter $r$, the above three-step LOCC
 protocols are close to optimum over all two-way LOCC protocols from the
 viewpoint of the Hoeffding bounds, too. 
Here, note that this is the first result showing the existence of a gap
between first error exponents under  optimal one-way LOCC protocol and under 
our two-way LOCC protocol in asymptotic settings of local discrimination problems.

\section{Summary}\label{sec summary}
In this paper, we have treated the local asymptotic hypothesis testing between
an arbitrary fixed bipartite pure state $\ket{\Psi}$ and the completely 
mixed state $\rho_{mix}$.
We have showed that the Stein's lemma type of
optimal error exponents are given as 
$\log d_A + \log d_B-E(\ket{\Psi})$ for all one-way LOCC, two-way LOCC and separable POVMs. 
The Chernoff bounds are given as $\log d_A + \log d_B -\log R_s(\ket{\Psi})$ for
one-way LOCC POVMs and as $\log d_A + \log d_B -LR(\ket{\Psi})$ for
separable POVMs. In
these formulas, $E(\ket{\Psi})$, $R_s(\ket{\Psi})$, and $LR(\ket{\Psi})$ are the entropy of entanglement,
the Schmidt rank, and the
logarithmic robustness of entanglement of $\ket{\Psi}$, respectively. 
From the viewpoint of the entanglement theory, these formulas give new
operational interpretations of these entanglement measures. 
Moreover, we have derived analytical formulas of the Hoeffding bounds for
one-way LOCC POVMs without any restriction on a parameter and  under separable
POVM with a restricted region of a parameter. Finally, we have numerically calculated 
the Chernoff and  Hoeffding bounds under a particular class of three-step LOCC protocols  in low-dimensional
      systems and have showed that these bounds not only
      outperform the bounds for one-way LOCC POVMs but also almost approximate the
      bounds for separable POVMs in the parameter region  where
      the analytical formula for separable POVMs has been derived. 
As far as we know, this is a first time to show the existence of a gap
between the optimal error exponent among one-way LOCC POVMs and two-way LOCC POVMs
in ``{\it asymptotic} local discrimination problems''.  

\section*{Acknowledgement}
MO thanks Dr. Go Kato for discussions. MO acknowledges 
support as a JSPS
Postdoctoral Fellow for Research Abroad.
This research was partially supported by the MEXT Grant-in-Aid for Young
Scientists (A) No. 20686026, and Grant-in-Aid for Scientific Research (A) No. 23246071. 
The Center for Quantum Technologies is funded by the Singapore
Ministry of Education and the National Research Foundation
as part of the Research Centres of Excellence programme.

\appendix
\section{List of notations}\label{sec notations}
\begin{description}
\item[$\overline{\lim}$:] limit superior
\item[$\underline{\lim}$] limit inferior
 \item[$\Hi_{A(B)}$:] Hilbert space of system $A(B)$
\item[$\Hi_{AB}$] Hilbert space of joint system $AB$
\item[$d_{A(B)}$:] dimension of $\Hi_{A(B)}$
\item[$d$:]$=\min\left(d_A,d_B\right)$
\item[$d_{max}$:]$=\max\left(d_A,d_B\right)$
\item[$\ket{\Psi}$:] an arbitrary pure state
\item[$\Psi$:]$=\ket{\Psi}\bra{\Psi}$
\item[$\sigma_A$:]$= \Tr_B \ket{\Psi}\bra{\Psi}$
\item[$\left\{\lambda_i\right\}_{i=1}^d$:] Schmidt coefficients of 
	    $\ket{\Psi}$ defined by Eq.~(\ref{eq schmidt decomposisiton psi})
\item[$\rho_{mix}$:] completely mixed state on $\Hi_{AB}$ defined by Eq.(\ref{eq def rho mix})
\item[$\sigma_\Psi$:] separable state defined by Eq.~(\ref{eq def sigma psi})
\item[$\ket{\psi}$:] the state on $\mathbb{C}^d$ defined by Eq.~(\ref{eq def ket psi})
\item[$\ket{\phi}$:] the state on $\mathbb{C}^d$ defined by Eq.~(\ref{eq def ket phi 0})
\item[$\left \{ \ket{\phi_{L}^n} \right \}_{L \in 
\mathbb{Z}^{d^n}_2}$:]  set of the states on 
	   $\left(\mathcal{C}^d\right)^{\otimes n}$ defined by Eq.~(\ref{eq def phi l n})
\item[$\{\ket{J_n} \}_{J_n=1}^{d^n}$:] abbreviation of the
basis $\{ \ket{i_1} \otimes \cdots \otimes \ket{i_n}\}_{i_1,\dots,i_n}$
\item[$\{T_n, I-T_n\}$:]two-valued POVM; $T_n$ corresponds to
	    $\rho_{mix}$ and $I-T_n$ corresponds to $\Psi$.
\item[$C$:] a class of POVM, e.g., $g$(global POVM), $\rightarrow$(one-way LOCC POVM), 
	   $\leftrightarrow$(two-way LOCC POVM), $sep$(separable POVM)
\item[$L_{\delta',n}^{A(B)}$:] projection defined by Eq.~(\ref{eq def L delta' n A})
\item[$\ket{\Upsilon_n}$:] state defined by Eq.~(\ref{eq def Upsilon n})
\item[$\mathcal{P}(d_A)$:] power set of a finite set $\{1,\dots , d_A\}$
\item[$\Lambda(t)$:] likelihood-ratio test defined by Eq.(\ref{eq def lrt})
\item[$\left\{M_{\omega}\right\}_{\omega \in \mathcal{P}(d_A)}$:] 
	   POVM of 
	   Alice's first measurement in the three-step LOCC protocol.
\item[$\{m_\omega(h)\}_{h=1}^{d_A}$:]positive coefficients of 
	   $M_{\omega}$ defined by Eq.~(\ref{eq def M omega})
\item[$\{\ket{\xi_j^\omega}\}_{j=1}^{|\omega|}$:]mutually unbiased 
	   basis of subspace $span\{\ket{h} \}_{h \in \omega}$ 
\item[$\{ N_j^\omega\}_{j=0}^{|\omega|}$:]POVM of Bob's measurement 
	   in the three-step LOCC protocol defined as $N_j^\omega=\ket{\xi_j^\omega}\bra{\xi_j^\omega}$ 
for $j \in \{1, \dots, |\omega| \}$, and $N_0^\omega \stackrel{\rm def}{=} I_B - \sum_{j=1}^{|\omega|}
      N_j^\omega$
\item[$\{
      O^{\omega j}_k \}_{k\in \{0,1\}}$:] POVM of Alice's second 
	   measurement in three-step LOCC protocol defined by Eq.~(\ref{eq def O omega j}) 
\item[$P(\delta|\rho_{in}, m_\omega)$:] classical probability 
	   distribution over measurement outcomes $\delta$
derived from the three-step LOCC protocol defined in Section \ref{sec two-way}, where $\rho_{in}$ is the 
	   unknown initial state, and Alice's first POVM $\{M_\omega 
	   \}_{\omega \in \mathcal{P}(d_A)}$ is determined by the 
	   parameter $\{m_\omega(h)\}_{h\in \omega}$. 
\item[$\alpha\left(T_n\right)$:]type-1 error probability of the local 
	   hypothesis testing defined by Eq.~(\ref{eq def type 1 error})
\item[$\beta\left(T_n \right)$:]type-2 error probability  of the local 
	   hypothesis testing defined by Eq.~(\ref{eq def type 2 error})
\item[$\alpha_{n,C}(\beta|\rho \| \sigma)$:] optimal type-1 error 
	    probability of POVM class $C$ defined by Eq.~(\ref{eq def alpha n C alpha rho sigma})
\item[$\beta_{n,C}(\alpha|\rho \| \sigma)$:] optimal type-2 error 
	    probability of POVM class $C$ defined by Eq.~(\ref{eq def beta n C alpha rho sigma})
\item[$P_{n,C}(\pi_0,\pi_1|\rho \| \sigma)$:] optimal mean error 
	    probability of  POVM class $C$ defined by Eq.~(\ref{eq def p n c pi 0 pi 1 rho sigma})
\item[$\xi_{C}(\rho\| \sigma)$:] Chernoff bound of POVM class $C$ 
	    defined Eq.~(\ref{eq def xi C})
\item[$\overline{\xi}_{C}(\rho\| \sigma)$:] upper bound of Chernoff 
	   bound $\xi_{C}(\rho\| \sigma)$ of POVM class $C$ 
	    defined by using $\overline{\lim}$ instead of $\lim$ in Eq.~(\ref{eq def xi C})
\item[$\underline{\xi}_{C}(\rho\| \sigma)$:] lower bound of Chernoff 
	   bound $\xi_{C}(\rho\| \sigma)$ of POVM class $C$ 
	    defined by using $\underline{\lim}$ instead of $\lim$ in Eq.~(\ref{eq def xi C})
\item[$\tilde{\xi}_{\leftrightarrow} \left(\rho_{mix} \| \Psi 
\right)$:] lower bound of Chernoff bound $\xi_{\leftrightarrow} \left(\rho_{mix} \| \Psi 
\right)$ defined by Eq.~(\ref{eq def tilde xi})
\item[$\eta_{C}(\epsilon|\rho\| \sigma)$:] Stein's lemma type of 
	   type-1 error exponent of POVM class $C$ 
	    defined by Eq.~(\ref{eq def stein alpha})
\item[$\overline{\eta}_{C}(\epsilon|\rho\| \sigma)$:] upper bound of 
 Stein's lemma type of 
	   type-1 error exponent $\eta_{C}(\epsilon|\rho\| \sigma)$ of POVM class $C$ 
	    defined by using $\overline{\lim}$ instead of $\lim$ in Eq.~(\ref{eq def stein alpha})
\item[$\underline{\eta}_{C}(\epsilon|\rho\| \sigma)$:] lower bound of 
Stein's lemma type of 
	   type-2 error exponent $\eta_{C}(\epsilon|\rho\| \sigma)$ of POVM class $C$ 
	    defined by using $\underline{\lim}$ instead of $\lim$ in Eq.~(\ref{eq def stein alpha})
\item[$\eta_{C}^\dagger(\rho\| \sigma)$:] strong converse bound of 
	   Stein's lemma type of  
	   type-1 error exponent of POVM class $C$ 
	    defined by Eq.~(\ref{eq def strong converse eta})
\item[$\widetilde{\eta}_{sep}(\rho\| \sigma)$:] upper bound of 
$\overline{\eta}_{sep}(\epsilon|\rho\| \sigma)$
	    defined by Eq.~(\ref{eq def tilde eta})
\item[$\theta_{C}(\epsilon|\rho\| \sigma)$:] Stein's lemma type of 
	   type-2 error exponent of POVM class $C$ 
	    defined by Eq.~(\ref{eq def stein beta})
\item[$\overline{\theta}_{C}(\epsilon|\rho\| \sigma)$:] upper bound of 
Stein's lemma type of 
	   type-2 error exponent $\theta_{C}(\epsilon|\rho\| \sigma)$ of POVM class $C$ 
	    defined by using $\overline{\lim}$ instead of $\lim$ in Eq.~(\ref{eq def stein beta})
\item[$\underline{\theta}_{C}(\epsilon|\rho\| \sigma)$:] lower bound of 
Stein's lemma type of 
	   type-2 error exponent $\theta_{C}(\epsilon|\rho\| \sigma)$ of POVM class $C$ 
	    defined by using $\underline{\lim}$ instead of $\lim$ in Eq.~(\ref{eq def stein beta})
\item[$\theta_{C}^\dagger(\rho\| \sigma)$:] strong converse bound of 
	   Stein's lemma type of  
	   type-2 error exponent of POVM class $C$ 
	    defined by Eq.~(\ref{eq def strong converse})
\item[$A_{C}(r|\rho\| \sigma)$:] Hoeffding bound of 
	   type-1 error exponent of POVM class $C$ 
	    defined by Eq.~(\ref{eq def a c})
\item[$\overline{A}_{sep}(r|\rho_{mix}\| \Psi)$:] upper bound of  
	   Hoeffding bound $A_{sep}(r|\rho_{mix}\| \Psi)$ 
	    defined Eq.~(\ref{eq def overline A sep})
\item[$\underline{A}_{sep}(r|\rho_{mix}\| \Psi)$:] lower bound of 
	   Hoeffding bound $A_{sep}(r|\rho_{mix}\| \Psi)$ 
	    defined by Eq.~(\ref{eq def underline A sep})
\item[$\tilde{A}_{\leftrightarrow} \left(r|\rho_{mix} \| \Psi 
\right)$:] lower bound of  Chernoff bound $A_{\leftrightarrow} \left(r|\rho_{mix} \| \Psi 
\right)$ defined by Eq.~(\ref{eq def tilde A})
\item[$B_{C}(r|\rho\| \sigma)$:]  Hoeffding bound of 
	   type-2 error exponent of POVM class $C$ 
	    defined Eq.~(\ref{eq def b c})
\item[$a_n\left(S_n\right)$:]  type-1 error probability of global 
	   hypothesis testing defined by Eq.~(\ref{eq def a n})
\item[$b_n\left(S_n\right)$:] type-2 error probability of global 
	   hypothesis testing defined by Eq.~(\ref{eq def b n})
\item[$\gamma_n \left( a \Big |\{\ket{\phi_{L}^n}\} \Big\| \ket{\psi} \right )$:]
optimal type-2 error probability of global hypothesis testing defined by Eq.~(\ref{eq def gamma n a})
\item[$\Gamma\left (r\Big |\{\ket{\phi_{L}^n}\} \Big\| \ket{\psi}\right 
 ) $:] Hoeffding bound of type-2 error exponent of global 
	   hypothesis testing defined by Eq.~(\ref{eq def gamma r big})
\item[$\Gamma_0(r)$] upper bound of  $\Gamma\left (r\Big |\{\ket{\phi_{L}^n}\} \Big\| \ket{\psi}\right 
 ) $ defined by Eq.~(\ref{eq def gamma  0 r})
\item[$Q_n(\kappa_0,\kappa_1)$:] optimal mean error probability of 
	   global hypothesis testing defined by Eq.~(\ref{eq def Q_n kappa _0 kappa_1})
\item[ $\kappa_{0(1)}(n)$:] prior probability defined by Eqs.~(\ref{eq def kappa 0}) and (\ref{eq def kappa 1})
\item[$D\left(\rho \|\sigma \right)$:] relative entropy defined by Eq.~(\ref{eq def relative entropy})
\item[$R_s(\ket{\Psi})$:] Schmidt rank of $\ket{\Psi}$ defined by Eq.~(\ref{eq def schmidt rank})
\item[$E\left(\ket{\Psi}\right)$:] the entropy of entanglement defined by Eq.~(\ref{eq def entropy of entanglement})
\item[$LR\left(\ket{\Psi}\right)$:] logarithmic robustness of 
	   entanglement defined by Eq.~(\ref{eq def LR})
\item[$f(s,{\bf m})$: ] the objective function defined by Eq.(\ref{eq two way f s omega = -log})
\item[$f(s)$: ] the objective function defined by Eq.(\ref{eq two way f s  = max})
\item[$s_r$:] the optimal parameter $s$ defined by Eq.(\ref{eq two way sr argmax})
\end{description}

\end{document}